\documentclass[review,sort&compress]{elsarticle}
\journal{Applied Energy}
\usepackage{graphicx,epstopdf,subfigure,booktabs,amssymb,amsmath}
\usepackage[dvipsnames]{xcolor}
\usepackage[hidelinks]{hyperref}

\newcommand{\kz}[1]{\multicolumn{1}{c}{#1}}

\newcommand{\new}[1]{\textcolor{black}{#1}}

\newcommand{\bi}{ \boldsymbol}
\usepackage{bm}

\begin{document}
\begin{frontmatter}

\title{Cell voltage model for Li-Bi liquid metal batteries}

\author[hzdr]{Norbert Weber\corref{cor1}}
\author[hzdr]{Carolina Duczek}
\author[hzdr]{Gerrit M. Horstmann}
\author[hzdr]{Steffen Landgraf}
\author[hzdr]{Michael Nimtz}
\author[hzdr]{Paolo Personnettaz}
\author[hzdr]{Tom Weier}
\author[mit]{Donald R. Sadoway}

\address[hzdr]{Helmholtz-Zentrum Dresden -- Rossendorf, Bautzner
  Landstr.\ 400, 01328 Dresden, Germany}
\address[mit]{Department of Materials Science and Engineering,
  Massachusetts Institute of Technology, 77 Massachusetts Avenue,
  Cambridge, MA 02139-4307, USA}

\cortext[cor1]{Corresponding author. Helmholtz-Zentrum Dresden –
  Rossendorf, Bautzner Landstr. 400, 01328 Dresden,
  Germany. \textit{E-mail address}: norbert.weber@hzdr.de (N. Weber)}

\begin{abstract}
Lithium-bismuth bimetallic cells are amongst the best explored liquid
metal batteries. A simple and fast \new{quasi-}one-dimensional cell voltage model
for such devices is presented. The equilibrium cell potential is obtained
from a complex two-dimensional fit of data drawn from multiple studies of equilibrium cell potential and rendered congruent with
the phase diagram. Likewise, several analytical and
fit functions for the ohmic potential drop across the electrolyte are
provided for different battery geometries. Mass transport
overpotentials originating from the alloying of Li into Bi are modelled by
solving a diffusion equation, either analytically or numerically, and
accounting for the volume change of the positive electrode. The
applicability and limitations of the model are finally illustrated 
in three distinct experimental settings.
\end{abstract}


%\tableofcontents

\end{frontmatter}

%\linenumbers

\section{Introduction}
\begin{figure}[h!]
\centering
\includegraphics[width=0.9\textwidth]{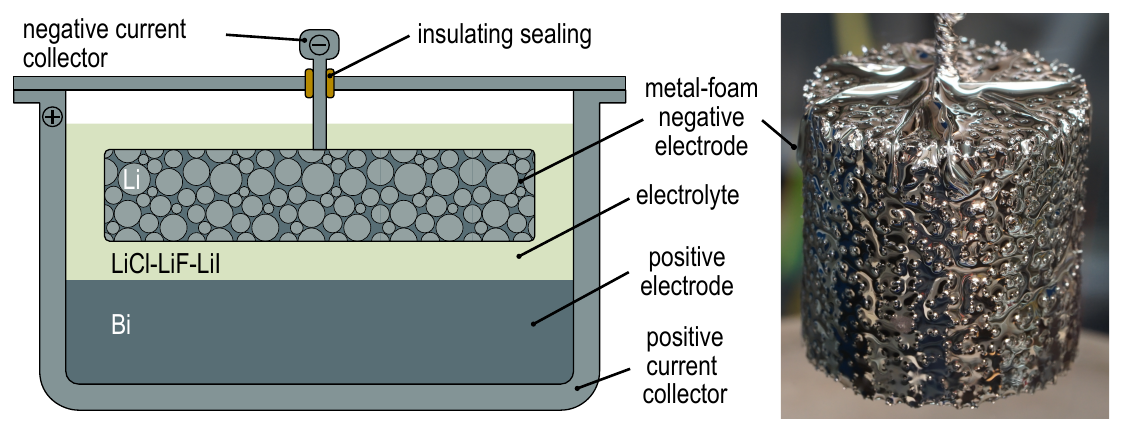}
\caption{Sketch (not to scale) of an LMB with a metal foam negative
  electrode (left) and Ni foam filled with Li (right).}
\label{f:setupLMB}
\end{figure}

The ever increasing deployment of highly fluctuating renewable
energies requires stationary energy storage to balance energy
production and consumption. Offering extreme current densities as well
as an extended lifetime at a competitive price, liquid metal batteries
(LMBs) have been discussed as an ideal candidate for large scale
energy storage \cite{Kim2013b}. While the heavy positive electrode metal
(e.g. Bi, cathode at discharge) forms the bottom layer, a
light negative electrode metal (e.g. Li, anode at discharge) is
usually soaked into a Ni foam -- as illustrated in
figure \ref{f:setupLMB}. Both liquid electrodes are separated by an 
ion-conducting (e.g. Li$^+$) molten salt mixture. At discharge, the
negative electrode metal is oxidised, crosses the electrolyte layer in
ionic form, is then reduced at the interface of the lower electrode
and finally alloys with the positive electrode metal.

Between the many possible electrode-combinations, the Li$||$Bi cell is
one of the best explored systems, as it provides unique benefits
compared to other chemistries. Table \ref{t:LiBicells} gives an
overview on the cells and their properties built in the past. On the
one hand, Li has a low solubility in its molten salts, which ensures a
very high coulombic efficiency by reducing self-discharge
\cite{Ning2015}. On the other hand, the Li$||$Bi cell potentially can
be operated in the two-phase area as well, which distinguishes it from
many other systems, as, e.g. Na$||$Bi or Ca$||$Bi
\cite{Ning2015}. When Li is alloyed into Bi during discharge, the
intermetallic phase Li$_3$Bi will form should a certain
Li-concentration be exceeded locally. The intermetallic phase would
then float on top of the Bi-electrode due to its lower density
\cite{Ning2015}. As Li$_3$Bi is an electric semiconductor and has a
very high diffusivity for Li \cite{Weppner1977a,Ning2015}, the cells
can be operated in a semi-solid state enabling a very high
Li-utilisation of more than 90\%. Although the intermetallic phase
dissolves slightly in molten salts, which is well visible due to its
red colour \cite{Foster1964a,Vogel1964,Vogel1964a} (see image in
\cite{Weier2017}), the solubility and irreversible Bi-transport to the
positive electrode is rather low
\cite{Vogel1965,Vogel1966,Ferris1973}.

\begin{table}
\caption{Properties of experimental Li$||$Bi cells: energy efficiency
  $\eta$, coulombic efficiency $\eta_\mathrm{c}$, operating
  temperature $T_\mathrm{o}$, maximum current density $j$, cell
  potential $E$ at a Li molar fraction $x_\mathrm{Li}$ and electrolyte
  layer thickness $h_\mathrm{E}$.}\label{t:LiBicells}
\centering
\tiny
\begin{tabular}{lrrrrrrrrrl}
\toprule
\kz{electrolyte} &
%\kz{$P$} &
%\kz{$E$} &
\kz{$\eta$} &
\kz{$\eta_\mathrm{c}$}&
\kz{life time}&
\kz{$T_\mathrm{o}$} &
\kz{price}&
\kz{$j$} &
\kz{$E$}&
\kz{$x_\mathrm{Li}$}&
\kz{$h_\mathrm{E}$}&
\kz{source}\\

\cmidrule(lr){2-3}
\cmidrule(lr){4-4}
\cmidrule(lr){5-5}
\cmidrule(lr){6-6}
\cmidrule(lr){7-7}
\cmidrule(lr){8-8}
%\cmidrule(lr){9-9}
\cmidrule(lr){10-10}
%\cmidrule(lr){11-11}

&%\kz{W/cm$^{2}$}&
%\kz{Wh/kg}&
\multicolumn{2}{c}{\%}&
\kz{cycles}&
\kz{$^\circ$C}&
\kz{\$/kWh}&
\kz{A/cm$^{2}$}&
\kz{V}&&
\kz{mm}\\
\midrule
 &    &  &  &   527   &  &  & 0.81 & 0.2 &  & \cite[p. 154]{Cairns1967} \\
 &    &  &  &527  &  &  &  0.9 &  &  & \cite[p. 186]{Vogel1966} \\
 &    &  &  &  &  &    & 0.931 &  &  & \cite[p. 119]{Lawroski1963b} \\
LiCl-KCl   &  &  &  &   489   &  &  & 0.93 &  & 15-33 & \cite[p. 155]{Cairns1967} \\
LiCl-KCl     &  &  &  &   489   && 0.8 & 0.93 & 0 &  & \cite[p. 113]{Chum1981} \\
LiCl-KCl   &  &  &  &   352   &  &  &  0.9 &  &  & \cite[p. 216]{Lawroski1963} \\
LiCl-KCl   &  &    &  & 400 &  & 1.6   & 0.94 &  &  & \cite{Temnogorova1979} \\
LiCl-KCl &&&&550&&6&&&&\cite{Nikitin1979}\\
LiCl-LiF  &  &  &  & 500-850 &  &  & 0.96 & 0.05 &  & \cite{Foster1964} \\
%LiCl-LiF &  &  &  &  &  & 435 &  &  & 0.74 & 0.75 &  & \cite[p. 109]{Cairns1967} \\
LiCl-LiF   &  &  &  & 500-850 &  &  & 0.96 & 0.05 &  & \cite[p. 109]{Cairns1967} \\
LiCl-LiF   & 70 & 99.7 & 1000 &   550 & 220-242 & 1.25   & 0.96 & 0 & 10 & \cite{Ning2015} \\
LiCl-LiF    &  &  &  & 500-850  &  &  &  0.96 & 0.05 &  & \cite{Lawroski1963a} \\
LiCl-LiF     &  &  &  & 477-880   &  &  & 0.96 & 0.05 &  & \cite[p. 118]{Lawroski1963b} \\
LiCl-LiF &&&&550&&6&&&&\cite{Nikitin1979}\\
LiCl-LiF &&&&510-560&&&0.96&0.03&&\cite{Demidov1973}\\
LiCl-LiF &&&&500-717&&&&&&\cite{Gasior1994}\\
LiCl-LiF-LiI  &  &  &  &   380-485 &  & 2.2   & 1.4 & 0 & 3.4 & \cite{Shimotake1969} \\
%LiCl-LiF-LiI &  &  &  & 100 &    &  &  & 1.8 &   0.8 &  & m3 & \cite[p. 54]{Shimotake1969} \\
%LiCl-LiF-LiI &  &  &  & 100 &    &  &  & 6.1 &   0.8 &  & m1 & \cite[p. 54]{Shimotake1969} \\
%LiCl-LiF-LiI  &  &  &  &   485   &  &  & 0.6 &  &  & \cite[p. 185]{Swinkels1971} \\
%LiCl-LiF-LiI   &  & 100 &  &  &  &  &  &  &  & \cite[p. 185]{Swinkels1971} \\
LiCl-LiF-LiI   &  &  &  &   485   && $>$1.8  & 1.25 & 0 & 3.4 &   \cite[p. 187]{Swinkels1971} \\
LiCl-LiF-LiI     &  &  &  &   500 &  & 2.2   & 0.9 &  &  & \cite[p. 111]{Cairns1973} \\
LiCl-LiF-LiI &&&&460&&&&&& \cite{Personnettaz2019}\\
\bottomrule
\end{tabular}\end{table}

Based on the good availability of material properties and experimental
data for Li$||$Bi cells, a number of modelling works have been
performed for the system \cite{Kelley2018}. Especially fluid dynamic
phenomena, such as the sloshing instability
\cite{Zikanov2015,Horstmann2017,Weber2017a}, electro-vortex flow
\cite{Weber2018}, thermal convection
\cite{Personnettaz2018a,Streb2018, Koellner2017}, Marangoni convection
\cite{Koellner2017}, mass transfer
\cite{Barriga2013,Ashour2018,Ashour2019} and solutal convection
\cite{Personnettaz2019,Personnettaz2020,Personnettaz2021} have been studied. In
addition, \new{one-dimensional electrochemical models for Mg-Sb \cite{Newhouse2014} and Li-Bi \cite{Newhouse2014,Personnettaz2019}
cells have been developed \cite{Newhouse2014},}
three-dimensional models for mass-transport overpotentials by various
authors \cite{Personnettaz2019,Herreman2020a} and a 3D cell voltage
model by Weber et al. \cite{Weber2019,Weber2020}. 

\new{All of these models have certain advantages and disadvantages:
  while the three-dimensional models are very accurate, they cannot
  be used for applications such as battery management systems as they
  are computationally too expensive. The one-dimensional models are
  fast, but sometimes ``oversimplified'', which limits their
  application to only certain use cases. The aim of the present paper
  is to discuss and classify the various effects determining the cell
  voltage of Li$||$Bi LMBs in order to develop a
  \new{quasi-}one-dimensional model, which unites the advantages of
  the 3D (accuracy) and 1D-models (speed). Being as simple as possible
  and as complex as necessary, it shall account for all relevant
  effects to be employed for a broad range of applications, as
  outlined in section \ref{s:useCases}.}

\section{Model}
\subsection{Overview}
The \new{quasi-}one-dimensional model describes the cell voltage as function of
current and time by first computing the equilibrium cell potential and
then subtracting the various overpotentials. Moreover, a diffusion
equation for the Li-concentration in Bi is solved in the positive electrode to
obtain (after a conversion) the Li molar fraction, which is needed to
calculate the equilibrium cell potential. Finally, volume changes of the
positive electrode and electrolyte layer are taken into account when solving the
diffusion equation and when determining the ohmic losses.

\subsection{Open circuit potential}\label{s:emf}
The equilibrium cell potential $E_\mathrm{eq}$ can be described by the Nernst equation
as \cite{Kim2013b}
\begin{equation}
E_\mathrm{eq} = -\frac{\mathrm{R}T}{\nu_e\mathrm{F}}\ln(a_\mathrm{Li(Bi)})
\end{equation}
with R denoting the universal gas constant, $T$ the temperature in
K, $\nu_e$ the number of exchanged electrons and F the Faraday
constant. Generally, the activity $a_\mathrm{Li(Bi)}$ of Li in Bi or $E_\mathrm{eq}$ itself
might be fitted, using e.g. data listed in table \ref{t:thermodynamics}. We take the latter option leading to a
molar fraction and temperature dependent equilibrium cell potential in
the \emph{liquid phase} of 
\begin{equation}\label{eqn:emf}
E_\mathrm{lq} = 10^{-3}(o + p\ln(x) + T  (a\ln(x) + bx + cx^2 + dx^3 + ex^4 + fx\ln(x)))
\end{equation}
with $T$ in K and
\begin{align}
o&=786.66,\quad
p=-6.10,\quad
a=-0.07,\quad
b=4.66,\\
c&=-16.50,\quad
d=28.96,\quad
e=-23.01,\quad
f=1.75.
\end{align}
The obtained fit function inter- and extrapolates the measurement
values of \cite{Foster1964,Demidov1973,Gasior1994}, and is valid for
temperatures between 415\,$^\circ$C and approximately 600\,$^\circ$C, with
higher errors up to 900\,$^\circ$C.
\begin{figure}[h!]
\centering
\includegraphics[width=0.9\textwidth]{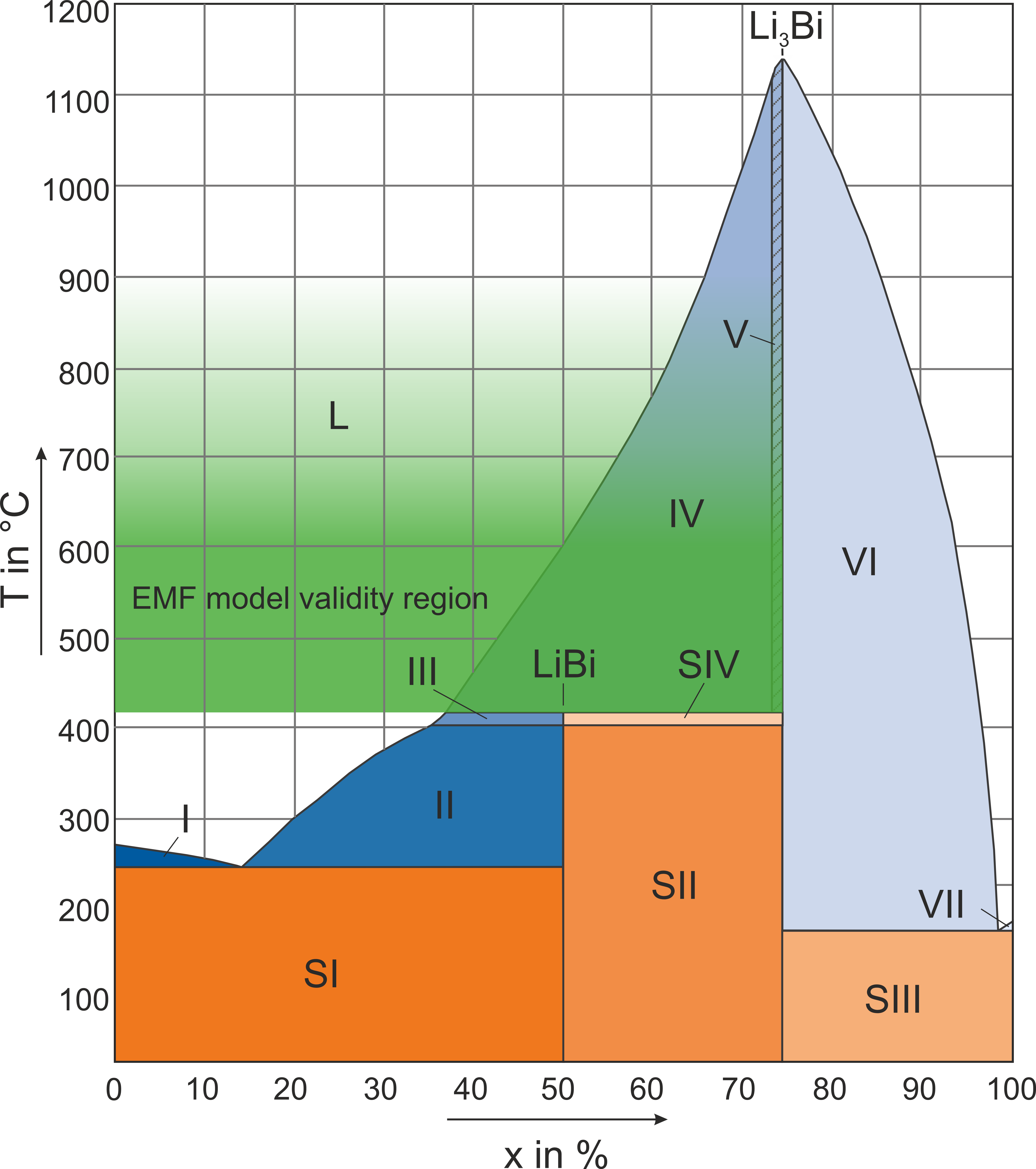}
\caption{Phase diagram of Li-Bi according to \cite{Predel1992} with
  different cell voltage areas and the region of validity shaded in
  green. \new{L denotes the liquid phase.}}\label{f:phaseDiagram}
\end{figure}

Once reaching the two-phase area (figure \ref{f:phaseDiagram}), the cell
potential will stay constant, when forming the
intermetallic phase Li$_3$Bi. The modelled equilibrium cell potential $E_\mathrm{eq}$ reads therefore
\begin{equation}\label{eqn:emfComplete}
E_\mathrm{eq}=\left\{\begin{array}{lll} 
E_\mathrm{lq} & \text{in the liquid area L,} \\
E_\mathrm{lq}(x=0.01) & \text{in the liquid phase L for x $<$ 0.01,} \\
E_\mathrm{lq}(T_\mathrm{lq}) & \text{in region IV,} \\
E_\mathrm{lq}(T_\mathrm{lq})\left(1 -\frac{x-0.73}{0.75-0.73}\right)
&\text{in region V.}\\
\end{array}\right.
\end{equation}
with $T_\mathrm{lq}$ denoting the liquidus temperature in K, obtained
from the phase diagram \cite{Sangster1991a} \new{using a spline fit.
The different phases and regions are listed in table \ref{t:phases}
and are identified in the model using simple if-else statements as
function of temperature and Li molar fraction.}
\begin{table}[h!]
	\caption{Phases present in the lithium-bismuth system. \new{L denotes the liquid phase.}}
	\label{t:phases}
	\centering
	\small
	\begin{tabular}{rrl}
		\toprule
		\multicolumn{1}{c}{region} & \multicolumn{1}{c}{phase}&state\\
		\midrule
		L & Li+Bi&liquid \\
		I & Bi+L&two-phase \\ 
		II & L+LiBi (low temp.)&two-phase \\
		III &  L+LiBi (high temp.)&two-phase \\
		IV &  L+Li$_3$Bi&two-phase\\
	    V & L+Li$_3$Bi (extended intermetallic)&two-phase \\
	    VI & Li$_3$Bi+L&two-phase\\
	    VII & L+Li&two-phase\\
	    SI & Bi+LiBi (low temp.)&solid solution \\
	    SII & LiBi (low temp.)+Li$_3$Bi&solid solution \\
	    SIII & Li$_3$Bi+Li&solid solution \\
	    SIV & LiBi (high temp.)+Li$_3$Bi&solid solution \\
		\bottomrule
\end{tabular}\end{table}

\begin{figure}[h!]
\centering
\includegraphics[width=0.7\textwidth]{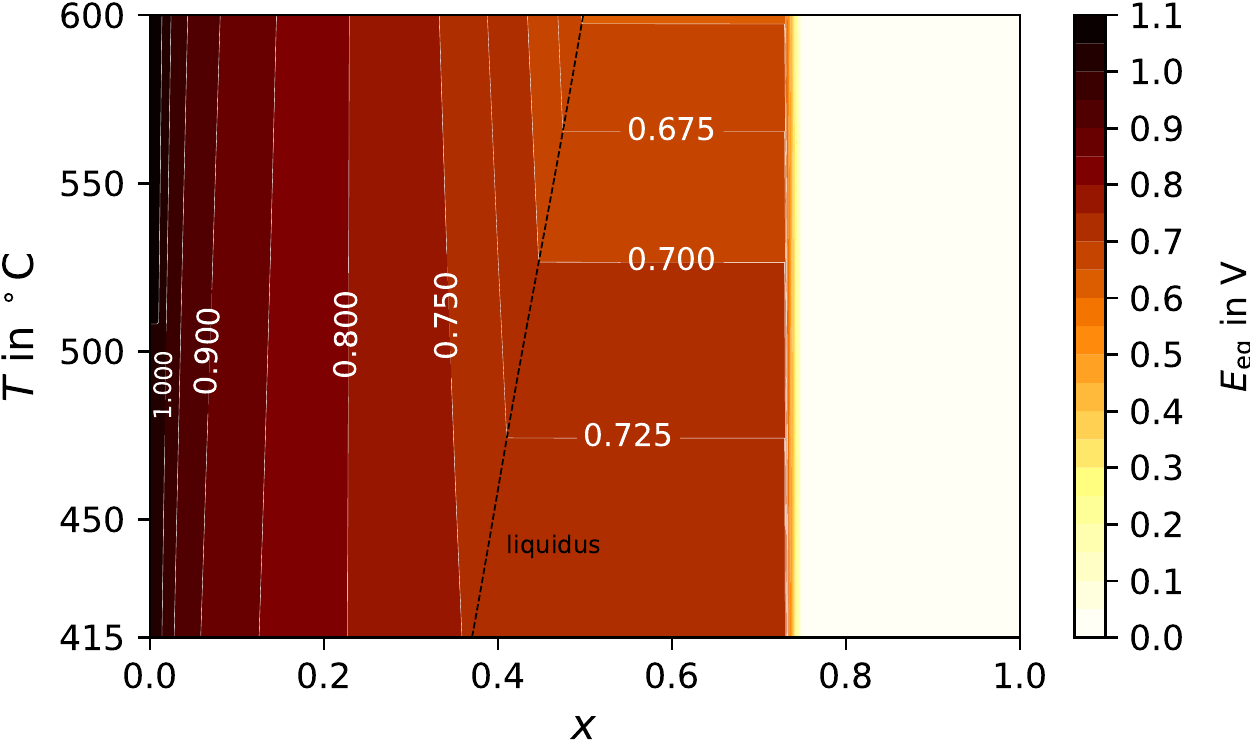}
\caption{Equilibrium cell potential fitted with experimental data from \cite{Foster1964,Gasior1994,Demidov1973}.}\label{f:emf}
\end{figure}

Regions not included in the cell voltage model are the phases with
high lithium fractions ($x>0.75$) and the liquid and solid
phases below 415\,$^\circ$C. While sufficient datapoints were available
for higher temperatures (450\,$^\circ$C to 900\,$^\circ$C), only few
datasets exist down to 380\,$^\circ$C for the two-phase region
\cite{Weppner1978}. The low-temperature liquid and bismuth-rich phases
are not represented in available measurement data. Therefore, we model
the electromotive force as follows in these phases:
\begin{equation}
E=0 \text{ in regions I-III, SI-SIV, VI and VII}.
\end{equation}
The resulting equilibrium cell potential as function of the Li molar fraction and temperature is
exemplarily illustrated in figure \ref{f:emf}.

\subsection{Overpotentials}
\subsubsection{Activation losses}
The activation overpotential $\eta_\mathrm{act}$ can be described by
the Butler-Volmer equation as \cite{Bard2001,Newman2004,Vetter1967}
\begin{equation}
j = j_0\cdot\left(\exp\left(\frac{\alpha n\mathrm{F}}{\mathrm{R}T}\eta_\mathrm{act}\right)-\exp\left(-\frac{(1-\alpha)n \mathrm{F}}{\mathrm{R}T}\eta_\mathrm{act}\right)\right)
\end{equation}
with $j$ denoting the current density, $j_0$ the exchange current
density, $\alpha$ the charge transfer coefficient and $n$ the reaction order of the rate limiting step. For a
small overpotential, i.e. if
\begin{equation}
\eta_\mathrm{act} < \mathrm{R}T / \mathrm{F} \approx 66\mathrm{mV}\quad(\mathrm{at}\ 500^\circ\mathrm{C}),
\end{equation}
the Butler-Volmer equation can be linearised by approximating
$\exp(y)$ as $1+y$, which leads to \cite{Bard2001}
\begin{equation}
j = j_0 \frac{n\mathrm{F}}{\mathrm{R}T}\eta_\mathrm{act}.
\end{equation}
Newhouse measured concentration dependent exchange current densities at the positive electrode between 4 and 50\,A/cm$^2$ \cite{Newhouse2014,Newhouse2017}. Using
both limiting values, we find for a typical temperature of 500\,$^\circ$C 
an activation loss of
\begin{equation}
\eta_\mathrm{act} = \left\{\begin{array}{ll}
0.0170\cdot j\ (\mathrm{A/cm}^2)\quad\mathrm{for}\quad j_0=4\mathrm{A/cm}^2\\
0.0013\cdot j\ (\mathrm{A/cm}^2)\quad\mathrm{for}\quad j_0=50\mathrm{A/cm}^2\end{array}\right. ,\\
\end{equation}
which would result in a voltage loss of $0.4\dots 5$\,mV for a typical
current density of 0.3\,A/cm$^2$ and to $1.3\dots 17$\,mV for a high
current density of 1\,A/cm$^2$. Assuming that the exchange current
densities are similar at the negative electrode, the activation losses appear to be
minuscule compared to the ohmic overpotential. In line with Newhouse'
conclusion \cite{Newhouse2017}, the activation losses are neglected in the present model.

\subsubsection{Ohmic overpotential}
The ohmic loss represents typically the most important overpotential
of an LMB. In general, its value can either be measured or
calculated. As the electrolyte conductivity is four orders of
magnitude lower than those of the metals, only ohmic losses in the
molten salt will be accounted for. In the most simple case, if
positive and negative electrode have the same cross section area $S$, the voltage loss is simply
\begin{equation}
\eta_\Omega = I R = I \frac{H}{S \sigma}
\end{equation}
with $I$ denoting the cell current, $R$ the electrolyte resistance,
$H$ its thickness, $\sigma$ the electrical conductivity and $S$ the
surface area. 

\begin{figure}[h!]
\centering
\includegraphics[width=0.7\textwidth]{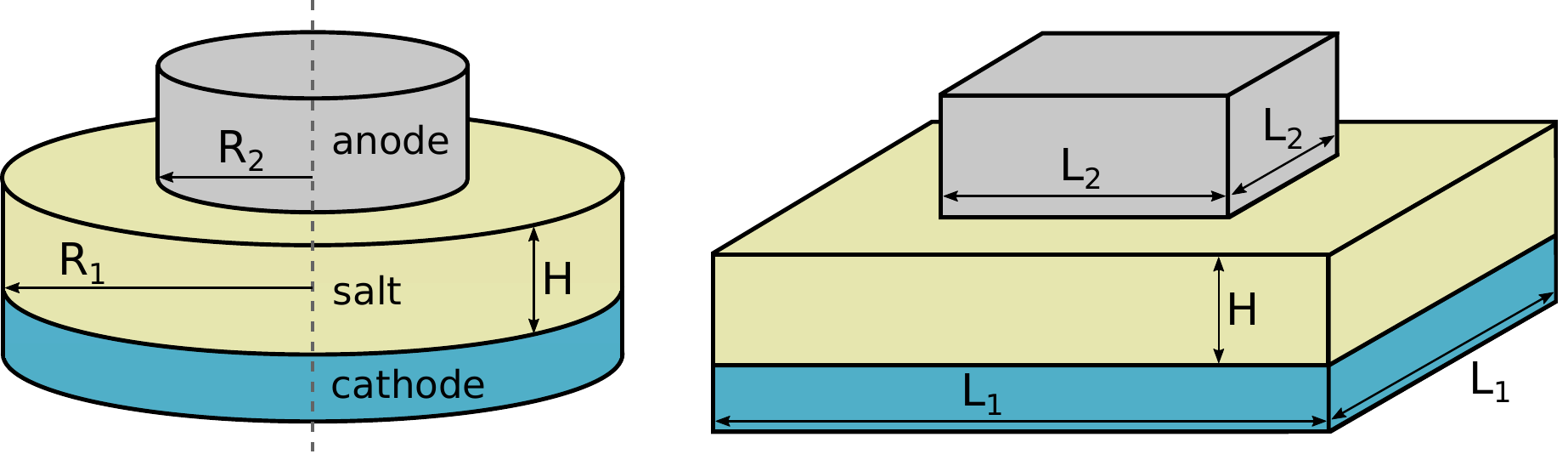}
\caption{Sketch of a cylindrical (left) and rectangular (right) LMB
using a metal-foam negative electrode current collector.}\label{f:foam}
\end{figure}

However, in real cells, often a metal foam is used to contain the
molten Li. As illustrated in figure \ref{f:foam}, the negative
electrode has then a smaller diameter than the positive electrode.
\new{Additionally, the porous structure of the foam may lead to an
  additional constriction resistance. However, as long as the pores at
  the interface are fully filled with molten Li and the ligaments are
  covered by it, i.e. as long as the cell is not fully discharged,
  such a constriction resistance might be neglected.}

For a cylindrical cell or square cell, as shown in
figure \ref{f:foam}, a simple Laplace equation can be solved in the
electrolyte  to obtain an estimate of the ohmic potential
drop. Considering the comparably low conductivity of the salt,
iso-potential boundary conditions are expected to provide the most
realistic results. Solving this equation for a rotational-symmetric
cylindrical geometry as well as a square cell with OpenFOAM
\cite{Weller1998} for a large parameter set and subsequent fitting, we
obtain the ohmic overpotential as function of current, radii and layer
height -- as outlined in \ref{a:ohmicCylinderFit} and
\ref{a:ohmicSquareFit}.

\begin{figure}[bh!]
\centering
\subfigure[]{\includegraphics[width=0.45\textwidth]{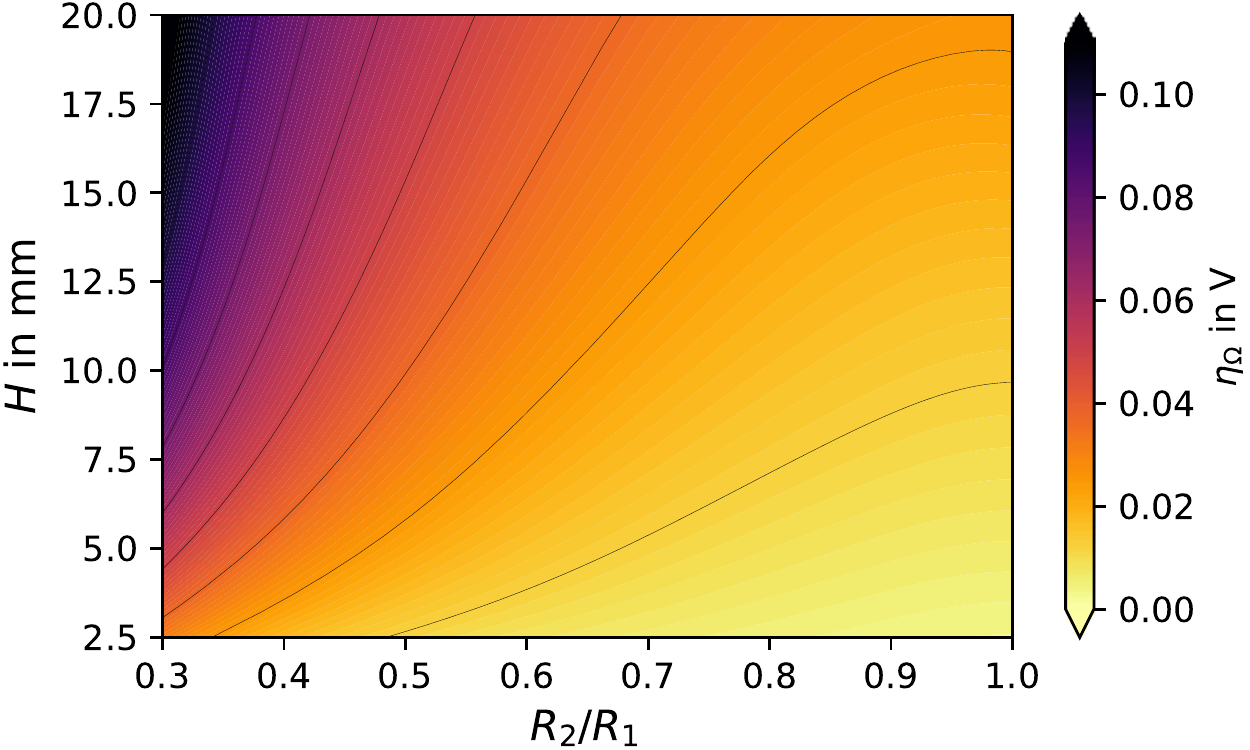}}
\subfigure[]{\includegraphics[width=0.45\textwidth]{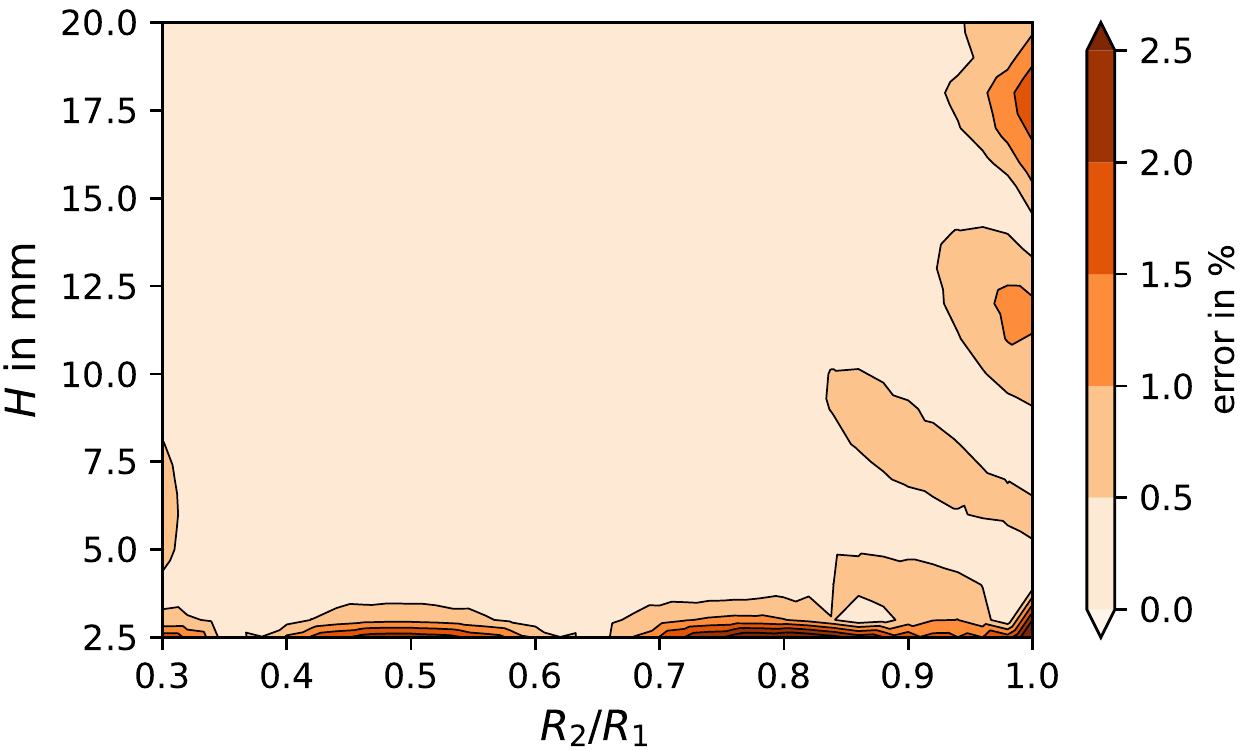}}
\subfigure[]{\includegraphics[width=0.45\textwidth]{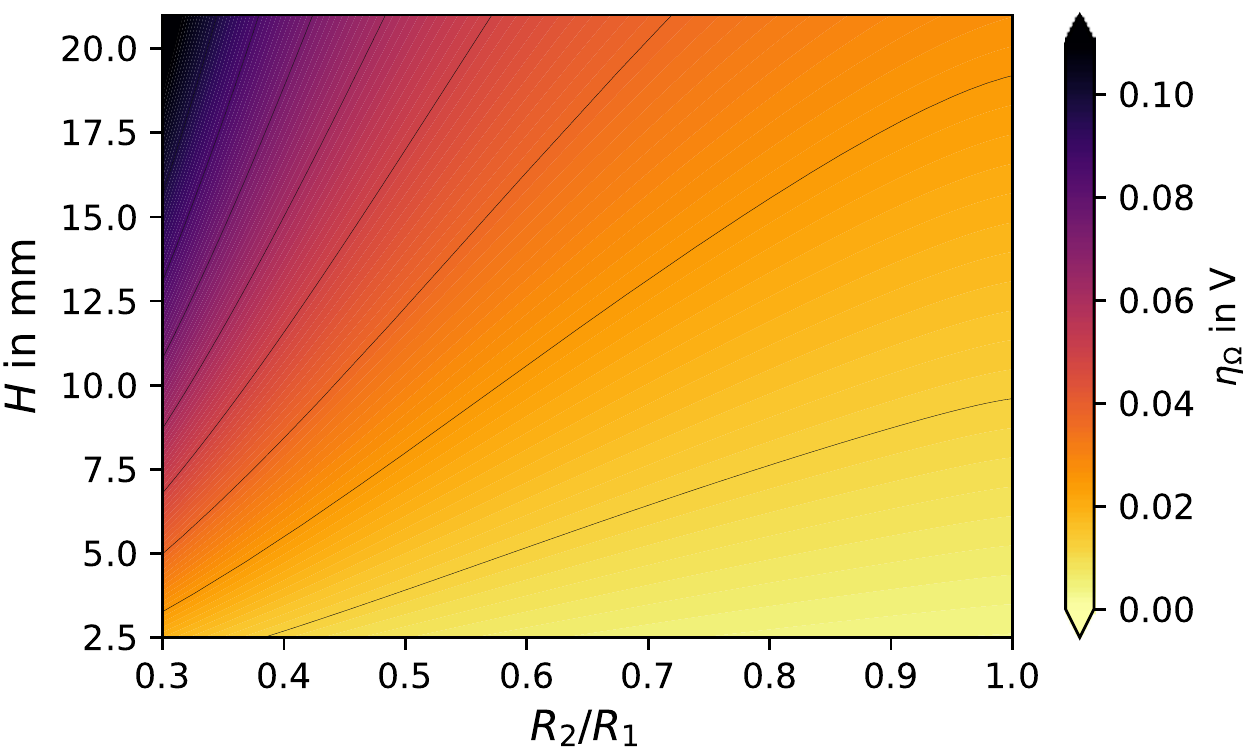}}
\subfigure[]{\includegraphics[width=0.45\textwidth]{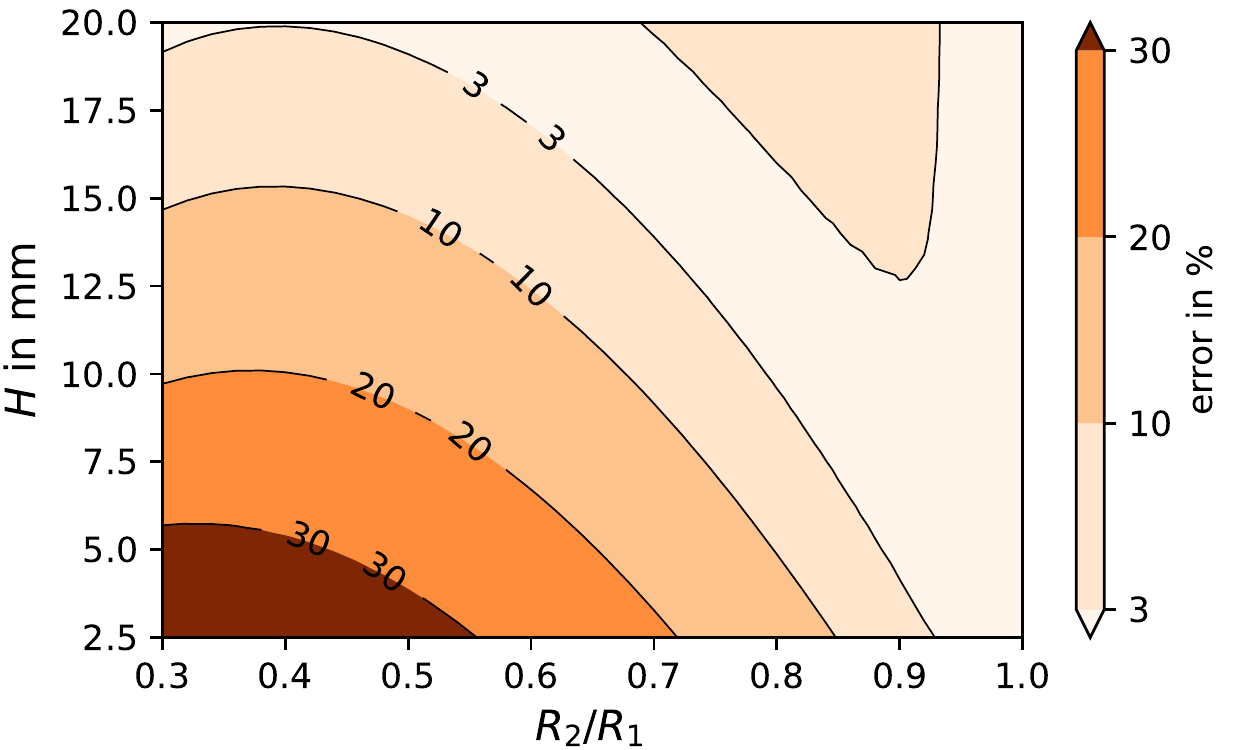}}
\subfigure[]{\includegraphics[width=0.45\textwidth]{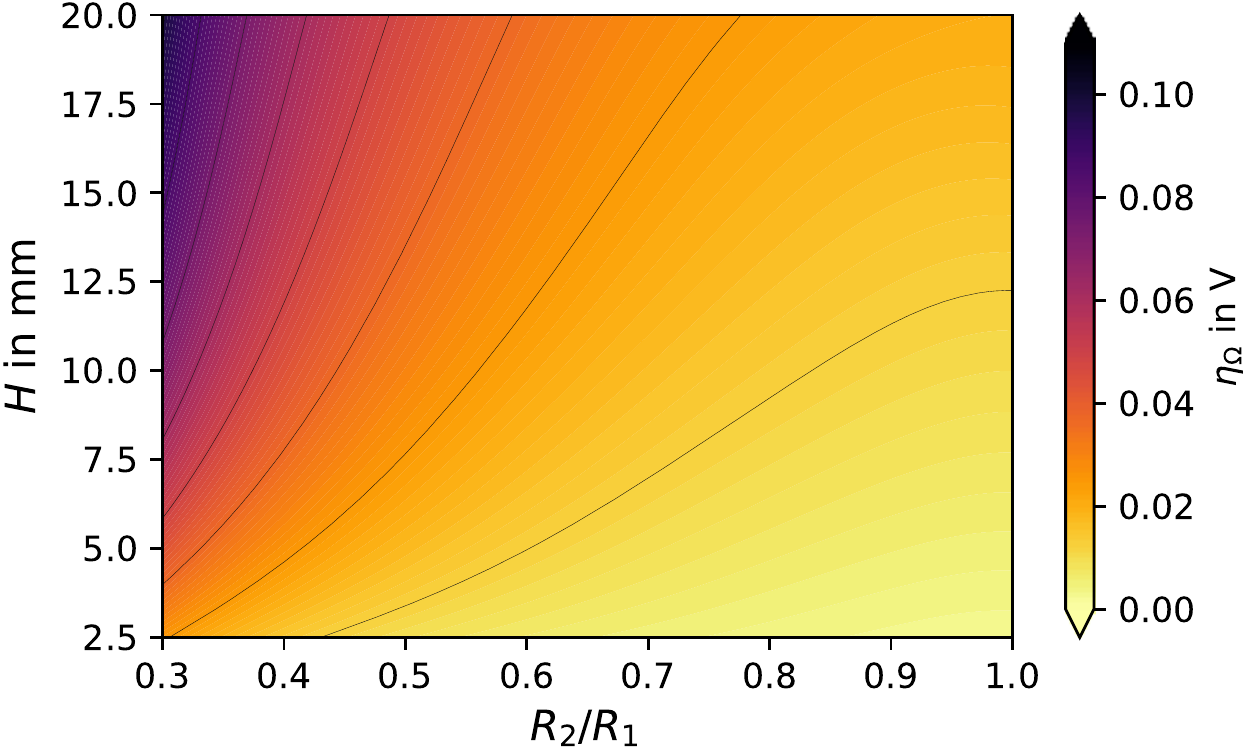}}
\subfigure[]{\includegraphics[width=0.45\textwidth]{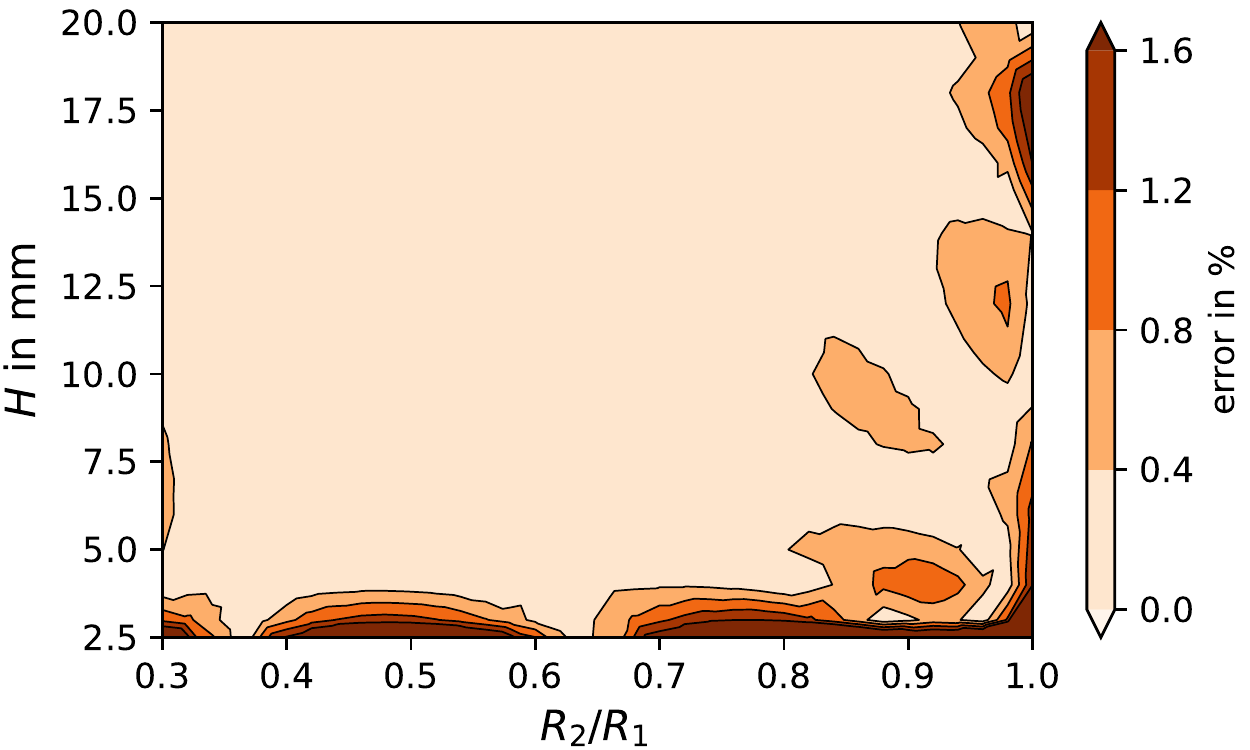}}
\caption{Fitted ohmic overpotential for a cylindrical cell (a) and
  corresponding error (b), analytically computed ohmic loss for the same cell (c) with
  error (d) and fitted ohmic overpotential for a square cell (e) with
  error (f). In the examples (a, c, e), $\sigma=1$\,S/cm, $I=1$\,A,
  $R_1=5$\,cm and $L_1=10$\,cm.}\label{f:ohmicLoss}
\end{figure}

Alternatively, the Laplace equation can be solved analytically for a
cylindrical cell using certain assumptions and simplifications -- as
explained in \ref{a:ohmAnalytical}. This leads to an explicit expression in terms of an infinite series:
\begin{align}
\eta_\Omega &= \frac{IH}{\sigma \pi R_1^2}\nonumber \\
&+ \sum_{n=1}^{\infty}\frac{IJ_1\left(\kappa_{0n}\right)}{\sigma \pi \epsilon_{0n}^2 R_2}\tanh\left(\frac{\epsilon_{0n}}{R_1}\frac{H}{2}\right) \frac{\pi H_0\left(\kappa_{0n}\right)J_1\left(\kappa_{0n}\right) + \left(2 - \pi H_1\left(\kappa_{0n}\right) \right)J_0\left(\kappa_{0n}\right)}{J_0^2\left(\epsilon_{0n}\right)}, \label{eq:Potential}
\end{align}
where $J_0$ and $J_1$ denote zero- and first-order Bessel functions of the first kind, $H_0$ and $H_1$ are zero- and first-order Struve functions and the numbers $\kappa_{0n}$ are given by $\kappa_{0n} 
= \epsilon_{0n}R_2 /R_1$, with $\epsilon_{0n}$ being the $n$ roots of the first derivative of the zero-order Bessel function, see \ref{a:ohmAnalytical} for more details. The solution appears to be rather intricate at first glance, but it allows for very fast calculations making it suitable for future integration into optimisation routines.  \\
Both the fitted and the analytical overpotential and corresponding relative errors are shown in figure \ref{f:ohmicLoss} for different heights of the salt layer $H$ and electrode sizes $R_2 / R_1$. The fitting functions lead to only very tiny errors, while the analytical formula deviates by up to $30\, \%$ from the OpenFOAM simulations for very small $H$ and $R_2$. This was to be expected because iso-current boundary conditions had to be applied for the solution (\ref{eq:Potential}), whereas complementary iso-potential boundary conditions were used in the simulation. However, the introduced error is fairly low in a large parameter range such that the analytical solution still provides a good and fast estimate of the voltage loss.

Finally, it should be noted that the salt layer thickness will vary
considerably during operation, if the negative electrode metal is contained in a
fixed metal-foam. This volume change needs to be accounted for as
explained in section \ref{s:volumeChange}.

\subsection{Diffusion equation}\label{s:diffusionEqn}
\subsubsection{General considerations}
The cell voltage of the LMB depends on the Li molar fraction at the
positive electrode-electrolyte interface. In order to obtain the latter, a
diffusion equation for the Li-concentration in Bi needs to be
solved. According to Newhouse \cite{Newhouse2014}, the finite
geometry of the positive electrode, its volume change and the concentration
dependent diffusion coefficient need to be taken into account.

The general diffusion equation reads \cite{Fick1855}
\begin{equation}\label{eqn:fick}
\frac{\partial c}{\partial t} = \nabla\cdot D \nabla c,
\end{equation}
with $D$ denoting the diffusion coefficient, $t$ the time and $c$ the
concentration. It can either be solved for the molar (mol/m$^3$) or
for the
mass concentration (kg/m$^3$). However, when including convection in
the cell voltage model, only the mass concentration is applicable any more,
because the Navier-Stokes equations are written for a mass-averaged
velocity. The density is strongly dependent on the concentration of Li in Bi. Therefore, the molar
and mass averaged velocities are not equal -- for a detailed
explanation, see Levicky \cite{Levicky2020}. Finally, it is not
admissible to solve directly for mass or molar fraction. The transport
equation of the latter is derived by dividing Fick's law
(equation \ref{eqn:fick}) by the density. As the density changes
continuously, when alloying Li into Bi, this operation is not permitted
\cite{Welty2015}. 

After solving the diffusion equation, the equilibrium cell potential can
be obtained by converting the concentration to molar fraction as
explained in section \ref{s:conversion}.

\subsubsection{Analytical solutions}
The diffusion equation (\ref{eqn:fick}) can be solved analytically for a
semi-infinite geometry with the concentration gradient $q$ at the electrolyte
interface as \cite[p.83]{Newhouse2014}
\begin{equation}
q = \nabla c\cdot\bi n = \frac{j}{\nu_e\mathrm{F}D},
\end{equation}
where $c$ denotes the molar concentration, $\bi n$ the normal vector and $\nu_e$ the number of
exchanged electrons. Using the approach of Carslaw
\cite[p. 75]{Carslaw1959} with $f = F_0 = Dq = j / (\nu_e\mathrm{F})$, $K=D$,
$\kappa = D$, we obtain the concentration at the interface as
\begin{equation}
c = c_0 + 2q\left(\frac{D t}{\pi}\right)^{1/2}
\end{equation}
with the initial concentration $c_0$. The concentration inside the
layer reads \cite{Carslaw1959} 
\begin{equation}
c = c_0 + 2q\left(\left(\frac{D t}{\pi}\right)^{1/2}\exp(-z^2/(4Dt))-\frac{z}{2}\mathrm{erfc}\frac{z}{2\sqrt{Dt}}\right),
\end{equation}
with $z$ denoting the vertical coordinate. By defining the integral of
the error function ierfc as
\begin{equation}
\text{ierfc}(z) = \frac{1}{\sqrt{\pi}}e^{-z^2} - z\cdot \text{erfc}(z)
\end{equation}
we obtain
\begin{equation}
c = c_0 + 2q\sqrt{Dt}\left(\text{ierfc}\frac{z}{2\sqrt{Dt}}\right),
\end{equation}
and replacing $z$ by $H-z$ with $H$ denoting the layer thickness finally leads to
\begin{equation}\label{eqn:diffInf}
c = c_0 + \frac{2j\sqrt{Dt}}{\nu_e\mathrm{F}D}\left(\text{ierfc}\frac{H-z}{2\sqrt{Dt}}\right).
\end{equation}
Here, the coordinate $z$ runs from the interface downwards. This
approach has already been used for LMBs by Personnettaz et
al. \cite{Personnettaz2019}. Of course, the simplification of a
semi-infinite layer is only valid for thick layers or short
charge/discharge times.

Alternatively, the diffusion equation can be solved analytically for a
finite layer with the same boundary condition at the electrolyte
interface as above, and $\nabla c \cdot\bi n = 0$ at the bottom
\cite{Newhouse2014}. Using the original solution for heat transfer
\cite[p. 112]{Carslaw1959} and setting $\rho c_p = 1$, $K=D$,
$\kappa = K/c_p = D$  and $F_0 = Dq = \frac{j}{\nu_e\mathrm{F}}$ we obtain
\cite[p. 61]{Crank1975}\cite{Herreman2020a}
\begin{equation}\label{eqn:diffFinite}
c = c_0 + \frac{jH}{\nu_e\mathrm{F}D} \left(\frac{Dt}{H^2}+\frac{3z^2-H^2}{6H^2}-\frac{2}{\pi^2}\sum_{i=1}^\infty\frac{(-1)^i}{i^2}\exp(-Di^2\pi^2t/H^2)\cos\frac{i\pi z}{H}\right).
\end{equation}
This approach has already been used for LMBs by Newhouse
\cite{Newhouse2014}; for alternative formulations of the same
equation, see \cite{Carslaw1959,Newhouse2014}.

\subsubsection{Numerical solution}
The analytical solutions of the diffusion equation are fast to
compute, but neither account for the concentration dependent
diffusion coefficient nor for volume change. Therefore,
equation (\ref{eqn:fick}) is discretised using finite differences and finite
volumes following Ferziger and Perić \cite{Ferziger2008} as explained
in detail in \ref{a:diffusionNumericalFDM} and
\ref{a:diffusionNumericalFVM}. The resulting equation system is solved
in Python \cite{VanRossum2009}.

Volume change is accounted for by adjusting the cell volume or
distance between the discretisation points in each time step. For this
purpose, the initial masses in each control volume (CV) or at each
point are computed, and then updated in every iteration. After
computing the density using equation (\ref{eqn:density:c}), the volume of
each CV is obtained. The latter is used to compute a correction
factor to increase either the size of the CV (finite volumes) or the
distance between the points (finite differences). As the amount of
moles of Li needs to stay constant during this operation, the
concentration needs to be divided by the same factor, when increasing
the volume.

\subsection{Conversion between concentration and molar fraction}\label{s:conversion}
The equilibrium cell potential is given as function of the molar fraction
(section \ref{s:emf}). However, it is not possible to solve for the
molar fraction directly -- instead, the equations need to be written
for molar or mass concentration (section \ref{s:diffusionEqn}). Hence,
a conversion between concentration and fraction is required in both
directions: from fraction to concentration to obtain the initial
condition for the diffusion equation (\ref{eqn:fick}) and from
concentration to molar fraction to obtain the cell voltage by equation 
(\ref{eqn:emf}).

The molar concentration of Li can be described as
\begin{equation}
c = \frac{n_\mathrm{Li}}{V} = \frac{x\cdot n}{V}
\end{equation}
with $n_\mathrm{Li}$ denoting the amount of Li in mol, $n$ denoting
the total amount of Li and Bi and $V$ the volume. This
gives with $n = m / M$, $V = m/\rho$ and $M = \sum_i x_i M_i$
\begin{align}\label{eqn:conversion}
c &= \frac{\rho(x) x}{xM_\mathrm{Li} + (1-x)M_\mathrm{Bi}},\\
x &= \frac{cM_\mathrm{Bi}}{\rho(c)-cM_\mathrm{Li}+cM_\mathrm{Bi}}.\label{eqn:conversion2}
\end{align}
As the density is usually only given as function of the molar fraction
$\rho(x)$ (section \ref{s:density}), the second formula cannot
simply be used. Converting concentration to molar fraction is possible
using one of the following simplifications:
\begin{enumerate}
\item Using a linear density law allows for an arbitrary conversion
  between concentration and fraction without further
  simplifications. As the linear approximation holds only near a
  working point, it should be used for short charge/discharge times
  only. A good domain of application are complex three-dimensional
  simulations \cite{Weber2020} together with the Oberbeck-Boussinesq
  approximation \cite{Oberbeck1879,Boussinesq1903}. 
\item The concentration of Bi might be assumed to be constant during
  the time of simulation. Then, the mole fraction is given as
  $x=c_\mathrm{Li} / (c_\mathrm{Li} + c_\mathrm{Bi})$. Again, this
  simplification should be limited to short discharge times.
\item The density might be fitted twice -- once for concentration and
  once for mole fraction, as explained in section
  \ref{s:density}. Even though both fits will not be perfectly
  equivalent, this allows an easy conversion between $c$ and $x$ and
  back using formulae \ref{eqn:conversion} and \ref{eqn:conversion2}.
\item Assuming the density to be constant is a very strong
  simplification, limited to very short operation times.
\item The equilibrium cell potential might be directly fitted as function
  of the concentration. Although this approach seems easy, as it
  eliminates the molar fraction from all equations, it has one
  drawback: the phase diagram needs to be converted from $x$ to
  $c$. In this step, it gets an additional dimension, because $c$
  depends on temperature, while $x$ does not.
\end{enumerate}
Within this work, we will use simplification (3) as it is the most accurate
one for long discharge times.

\subsection{Volume change}\label{s:volumeChange}
When the cell is discharged, and Li alloys into Bi, the positive electrode
thickness will increase. This volume change needs to be accounted for,
when solving the diffusion equation \cite{Newhouse2014}. In most of
the LMBs built in the past, the Lithium was contained in a metal-foam
current collector, mounted at a fixed position. If in these cells the
positive electrode layer changes its thickness, the electrolyte layer height will
change at the same time. Considering the high ohmic resistance of the
molten salt, the volume change of the electrolyte might be a very
important effect.

\begin{figure}[bth]
\centering
\includegraphics[width=0.5\textwidth]{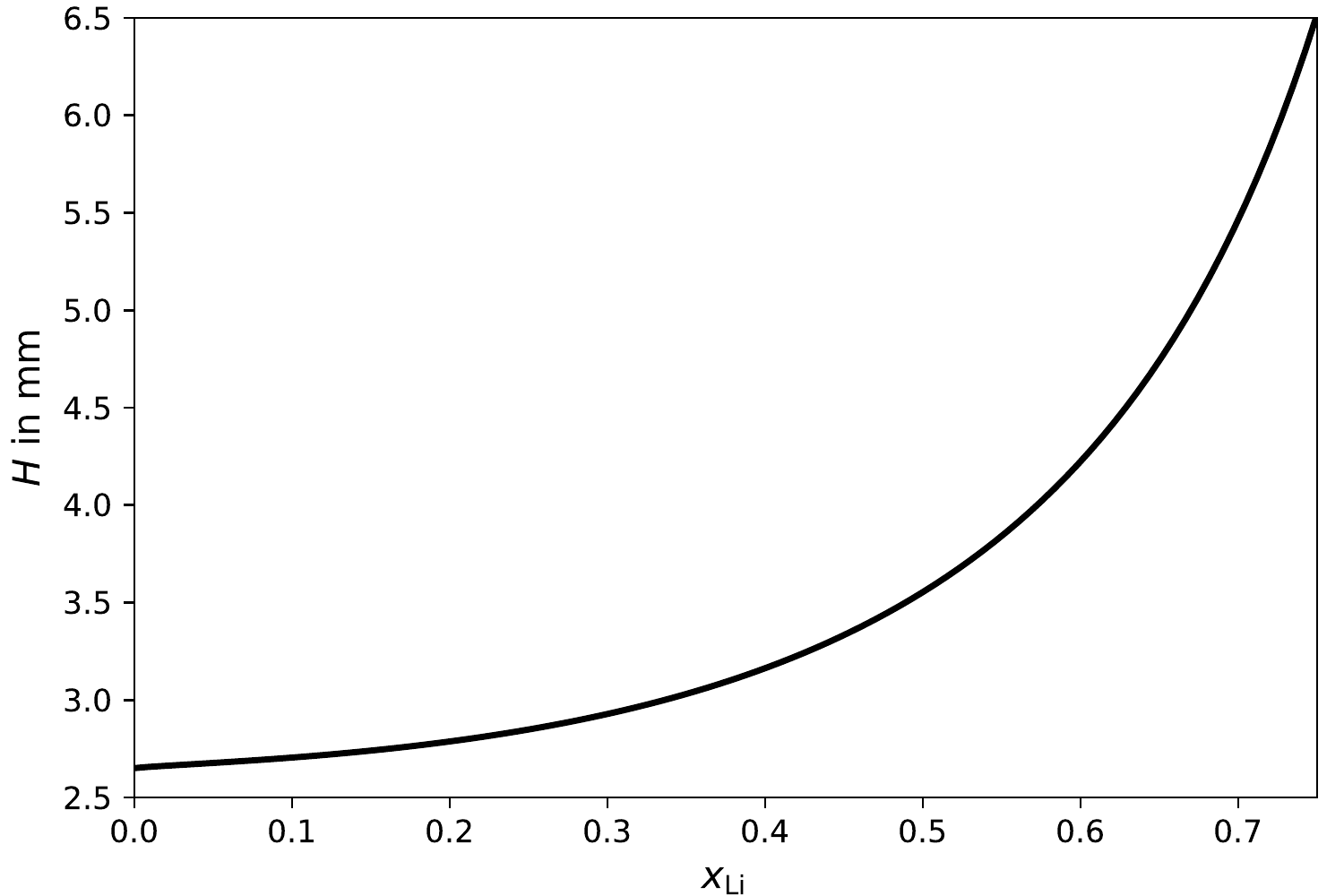}
\caption{Thickness of the positive electrode layer as function of the Li molar
  fraction in Bi for the experiment described in section
  \ref{s:Ning}.}\label{f:volumeChange}
\end{figure}

The change of height of the positive electrode layer during operation can be estimated as
\begin{equation}
\Delta H =
\dfrac{x\frac{n_\mathrm{Bi}}{1-x}M_\mathrm{Li}+m_\mathrm{Bi}}{S\rho(x)}
- \dfrac{x_0\frac{n_\mathrm{Bi}}{1-x_0}M_\mathrm{Li}+m_\mathrm{Bi}}{S\rho(x_0)}
\end{equation}
with $x_0$ denoting the initial Li molar fraction, $m$ the mass and $S$ the
surface area. Figure \ref{f:volumeChange} gives an illustrative example
for the large Li-Bi cell experiment, described in section
\ref{s:Ning}. There, the positive electrode layer changes between charged and
discharged state between a thickness of 2.7 and 6.5\,mm, i.e. roughly
by a factor of two. The volume change is considerable -- and needs to
be accounted for when solving the diffusion equation, but also when
computing the ohmic overpotential.

An elegant way to account for volume change in LMBs when solving the
diffusion equation has been introduced by Newhouse \cite{Newhouse2014}. 
She solved the diffusion equation in a
solvent-fixed reference frame, which means that no net flow of Bi (the
solvent) occurs over the single control volumes
\cite{Kirkwood1960,Wendt1959,Dufreche2002}. This approach involves
introducing a new spatial unit $\xi$, which describes the distance
between two solvent atoms, and to convert the diffusion
coefficient to this reference frame. After solving the modified
diffusion equation, and obtaining the Li-flux relative to the
Bi-atoms, the solvent-fixed concentration needs to be reconverted to
the ``true'' concentration. This approach is based on the original
article by Hartley and Crank \cite{Hartley1949} (available also in
\cite{Crank1975}), which is very well explained by Gekas and Lamberg
\cite{Gekas1991} and has been used for other applications, as well
\cite{Viollaz1984,Alsoy2002}. Apart from the difficult derivation, the
biggest drawback of this approach is the assumption of constant
partial volumes -- i.e. the assumption that the alloy density can be
described as that of an ideal solution (equation \ref{eqn:idealSolution}).

An alternative approach is to compute every single cell volume in each 
time step, when solving the diffusion equation. Then, the size of
the control volume can be updated in real-time. When solving the
diffusion equation in 1D, as explained in
\ref{a:diffusionNumericalFDM} and \ref{a:diffusionNumericalFVM}, the
distances between two nodes need to be scaled,
respectively.

\section{Material properties}
\subsection{Conductivity}
The electrical conductivity of the molten salt is required to compute
the ohmic losses. In the past, mostly mixtures of LiCl-LiF
\cite{Foster1964,Cairns1967,Ning2015,Lawroski1963a,Lawroski1963b} and
LiCl-LiF-LiI \cite{Shimotake1969,Swinkels1971,Cairns1973} have been
used for Li-Bi cells. LiCl-KCl was employed as well
\cite{Lawroski1963,Cairns1967,Temnogorova1979,Chum1981}, although it
is known that Li reduces KCl at temperatures above 500\,$^\circ$C
\cite{Foster1964,Blanchard2013}. Table \ref{t:conductivity} gives an
overview on the melting temperature and conductivity of the salts.

\begin{table}[h!]
\caption{Electric conductivity $\sigma$ and melting temperature
  $\vartheta_m$ of the molten salts.}
\label{t:conductivity}
\centering
\tiny
\begin{tabular}{lrrrl}
\toprule
\multicolumn{1}{c}{salt} & \multicolumn{1}{c}{composition in mol\%} & \multicolumn{1}{c}{$\vartheta_m$ in $^\circ$C} &  \multicolumn{1}{c}{$\sigma$ in S/cm} & \multicolumn{1}{c}{source}\\
\midrule
\textbf{LiCl-LiF} & 70:30&501&&\cite{Ning2015}\\
&&&6.78 at 927\,$^\circ$C&\cite{Redkin2011}\\
&various&&&\cite{Janz1979}\\
\midrule
\textbf{LiCl-LiF-LiI}&eutectic&340.9&3 at 475\,$^\circ$C& \cite{Shimotake1969}\\
&eutectic&340.9&2.3 at 475\,$^\circ$C&\cite[p. 97]{Vogel1968b}\\
& 29.1:11.7:59.2&340.9&2.3 at 375\,$^\circ$C&\cite{Shimotake1969}\\
&29.1:11.7:59.2&341&2.3 at 375\,$^\circ$C&\cite[p. 167]{Swinkels1971}\\
&29.1:11.7:59.1&341&8.895 exp(-872.6/$T$)& \cite{Masset2006b}\\
\midrule
\textbf{LiCl-KCl} && 354 & 1.83 at 500\,$^\circ$C & \cite{Fujiwara2010}\\
&40.45:59.55 && $13.21  \exp(-13995.71415 / RT)$ & \cite[p. 205]{Janz1988}\\
&41:59 & 353 & 1.7 at 476\,$^\circ$C& \cite{Kim2013b}\\
&58.8:41.2 & 353 & 1.57 at 450\,$^\circ$C& \cite[p. 167]{Swinkels1971}\\
&58.8:42.2 & & $18.7876 \exp(-1800.6/T)$ & \cite{Masset2006b}\\
&58.8:41.2 &&  $23.021 \exp(-16204.90311 / RT)$ & \cite[p. 205]{Janz1988}\\
&70.36:29.64& \multicolumn{2}{c}{$-11.0108 + 2.7461\cdot 10^{-2} T -13.2471\cdot10^{-6} T^2$} & \cite[p. 1045]{Janz1975}\\
&81.77:18.23&& $13.886 \exp(-9982.35421 / RT)$ & \cite[p. 205]{Janz1988}\\
&various &&& \cite[p. 1045]{Janz1975}\\
&various &&& \cite{vanArtsdalen1955}\\
\bottomrule
\end{tabular}\end{table}

\subsection{Density}\label{s:density}
\begin{figure}[h!]
\centering
\subfigure[]{\includegraphics[width=0.45\textwidth]{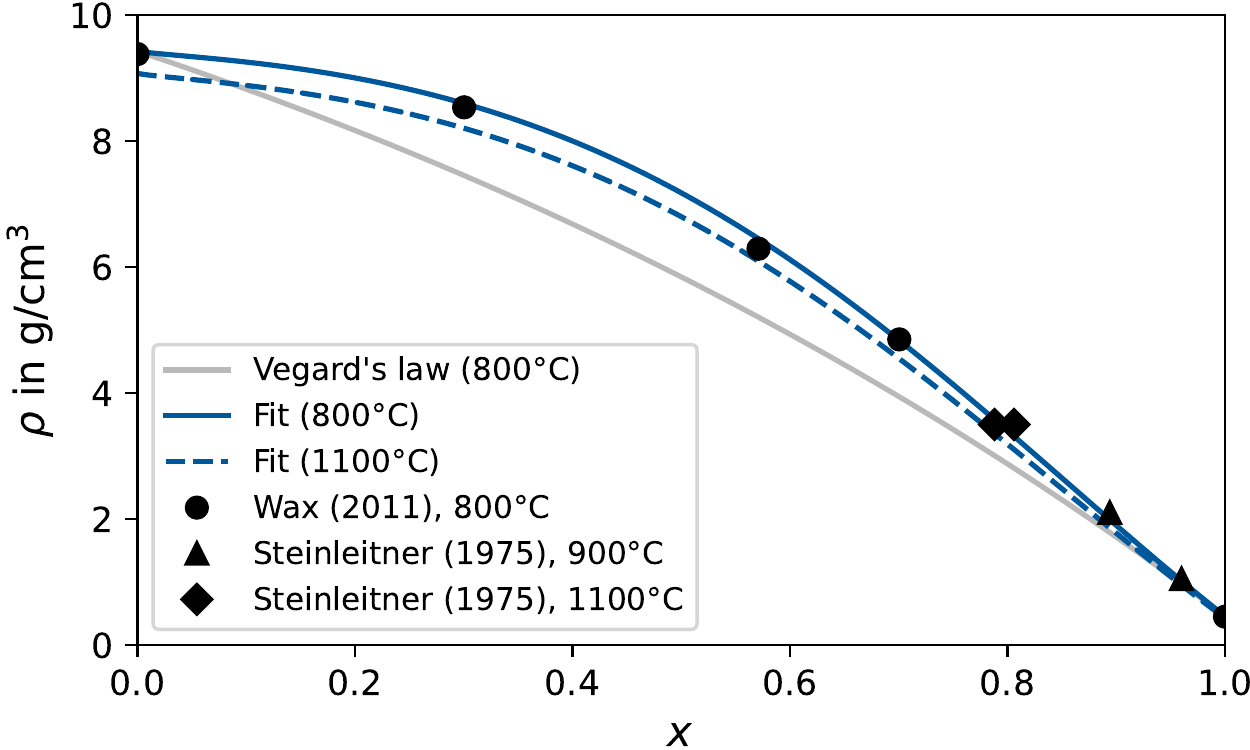}}
\subfigure[]{\includegraphics[width=0.45\textwidth]{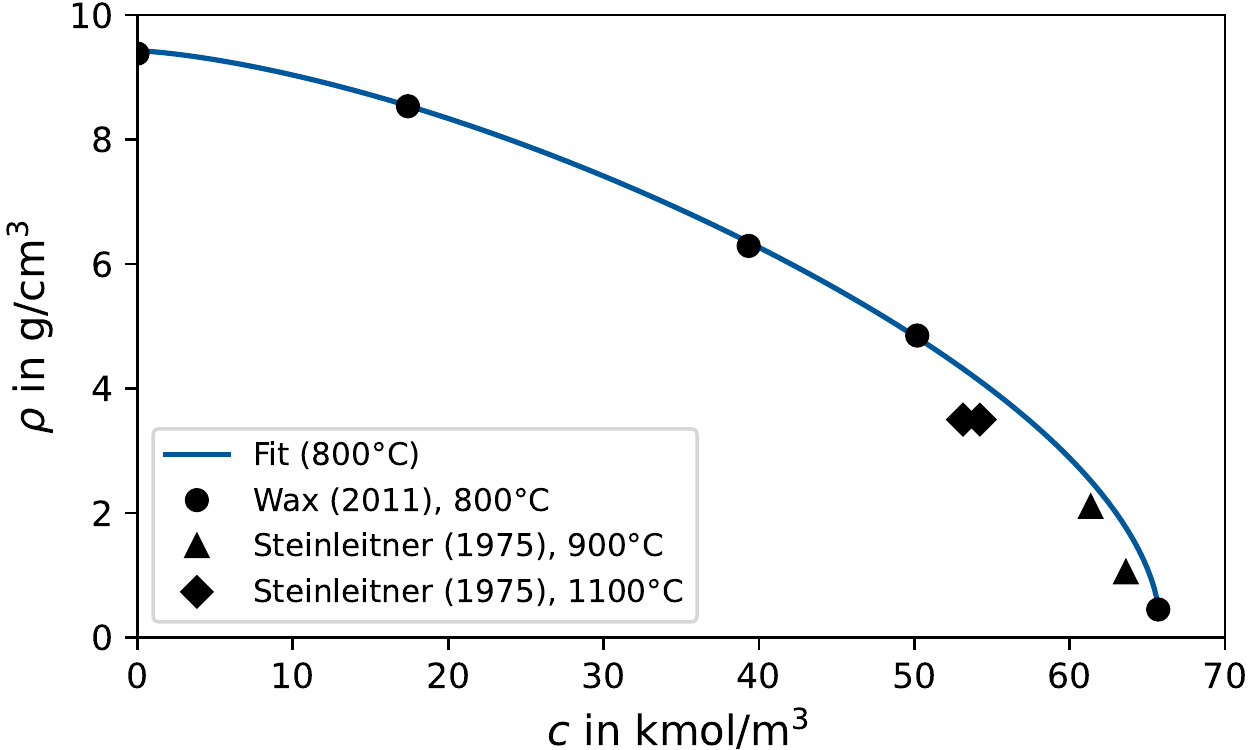}}
\caption{Density of Li-Bi depending on the Li molar fraction (a) and Li concentration (b).}\label{f:rho}
\end{figure}

The densities of molten Li and Bi are given in kg/m$^3$ as
\cite[p. 14-10]{Gale2004}
\begin{align}
\rho_\mathrm{Li} &= 518 - 0.1(T-453.5)\\
\rho_\mathrm{Bi} &= 10050 - 1.18(T-544)
\end{align}
with $T$ denoting the temperature in K. Densities of binary alloys are
sometimes estimated based on Vegard's law \cite{Vegard1921}. The
latter predicts a linear relation of the mean lattice distance in a
solid solution, if two different components are mixed. Applied to
fluid mixtures, it is assumed that the volume $V$ can be computed as a
linear weight of the amounts of the substances $n_i$ and the third
roots of their molar volumes $v_i$ as \cite{Fazio2015} 
\begin{equation}
V = \sum\left(n_i\sqrt[3]{v_i}\right)^3,
\end{equation}
which leads for a binary alloy of component 1 and 2 to \cite{Fazio2015}
\begin{align}
\rho = \frac{x_1M_1 + (1-x_1)M_2}{\left(x_1\sqrt[3]{M_1/\rho_1} + (1-x_1)\sqrt[3]{M_2/\rho_2}\right)^3}.\label{eqn:Vegard3}
\end{align}
Alternatively, the volume of an \emph{ideal solution} can be described
as the sum of its components as
\begin{equation}
V = \sum n_iv_i,
\end{equation}
which leads to the very similar equation \cite{Brillo2016}
\begin{equation}\label{eqn:idealSolution}
\rho = \frac{x_1M_1 + (1-x_1)M_2}{\left(x_1M_1/\rho_1 + (1-x_1)M_2/\rho_2\right)}.
\end{equation}
Both density laws are not well suited for Li-Bi, as can be seen in
figure \ref{f:rho}.
% While Vegard's law assumes the substitution of
%single atoms in a solid solution, the very different size of the Li-
%and Bi-atoms would possibly rather favour an interstitial solid
%solution -- where Vegard's law might fail. The strong difference in
%size of both atoms might similarly explain why Li-Bi is not an ideal
%solution.
The deviation from Vegard's law perhaps can be explained by electronic structure. While Li is strongly electropositive ($\chi=0.98$), Bi is considerably more electronegative ($\chi=1.9$). This difference might cause a deviation from ideality for the liquid solution -- similar as the Hume-Rothery Rules predict for solid state solutions a deviation from ideality for large electronegativity differences.

Only few density values for the Li-Bi alloy are available in the
literature
\cite{Steinleitner1975,Wax2011a,Hafner1985a,Souto2013,Bove2007}. These
data are used to fit the density as
\begin{equation} \label{eqn:density:x}
\rho = \frac{xM_\mathrm{Li} + (1-x)M_\mathrm{Bi}}{\left(x^{0.94}M_\mathrm{Li}/\rho_\mathrm{Li} + (1-x)^{1.64}M_\mathrm{Bi}/\rho_\mathrm{Bi}\right)}.
\end{equation}
As illustrated in figure \ref{f:rho}, this fitted function is describing
the Li-Bi density best, and is therefore used in the following.

In order to allow an easy conversion between concentrations and molar
fractions, the density is fitted as function of the concentration, as
well. For this purpose, the molar fractions belonging to the measured
density values are first converted to the Li molar concentrations as
\begin{equation}\label{eqn:c}
c = \frac{\rho x}{xM_\mathrm{Li} + (1-x)M_\mathrm{Bi}}
\end{equation}
with $\rho$ denoting the mixture density. The
resulting fit function fulfils the boundary conditions
\begin{align}
\rho(c=0) &= \rho_\mathrm{Bi}\\
\rho(c=c_\infty) &= \rho_\mathrm{Li}
\end{align}
and reads
\begin{equation} \label{eqn:density:c}
\rho = \frac{\alpha\cdot c^\gamma + \beta\cdot(c_\infty-c)^\delta}{\alpha\cdot\frac{c_\infty^\gamma}{\rho_\mathrm{Li}}\left(\frac{c}{c_\infty}\right)^\varepsilon + \beta\cdot\frac{c_\infty^\delta}{\rho_\mathrm{Bi}}\left(\frac{c_\infty-c}{c_\infty}\right)^\varphi}
\end{equation}
with the concentration of pure Li defined as
\begin{equation}
c_\infty = \frac{\rho(x=1)}{M_\mathrm{Li}} = \frac{\rho_\mathrm{Li}}{M_\mathrm{Li}}
\end{equation}
and
\begin{equation}
\varepsilon=1.45,\: \varphi=0.67,\: \alpha=8.1\cdot10^{-9},\: \beta=37,\: \gamma=2.37\ \mathrm{and}\ \delta= 0.7.
\end{equation}

\subsection{Diffusion coefficient}
\begin{figure}[h]
\centering
\includegraphics[width=0.7\textwidth]{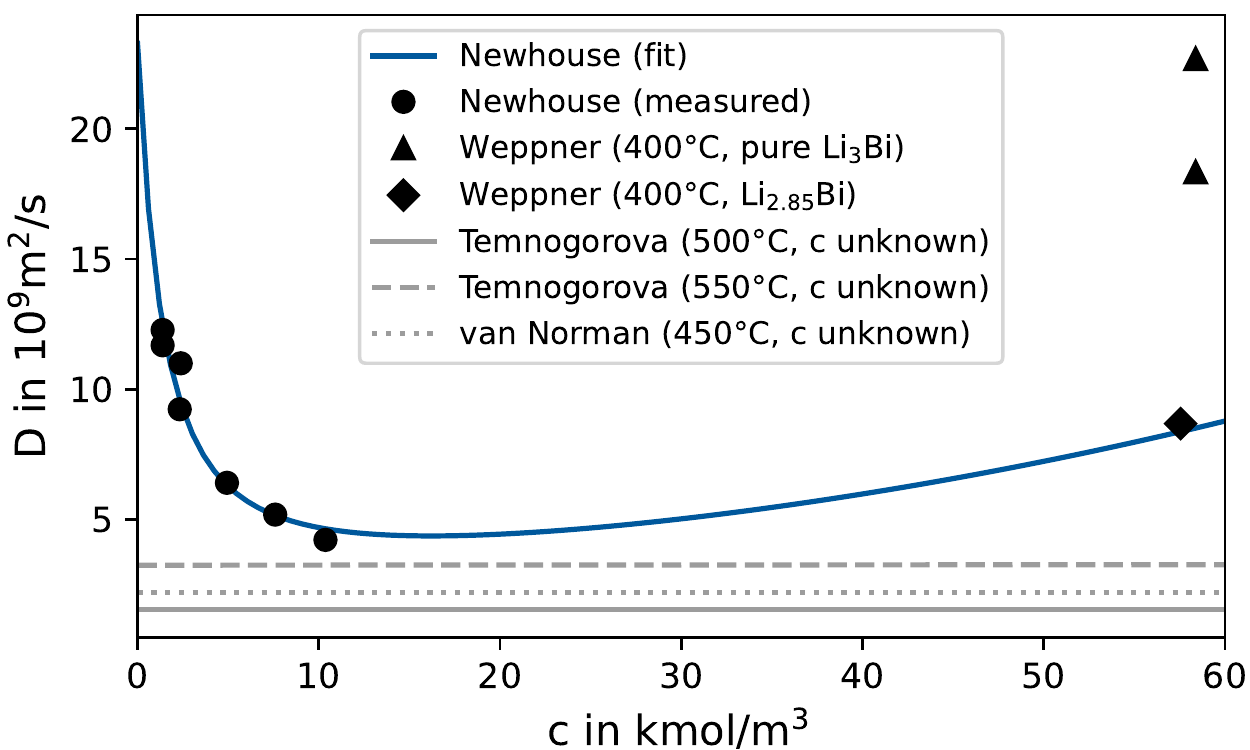}
\caption{Concentration dependent diffusivity according to various
  authors and fit by Newhouse \cite{Newhouse2014}; $x$ has been
  converted to $c$ using equations (\ref{eqn:density:x}) and
  (\ref{eqn:c}).}\label{f:diffusivity} 
\end{figure}

Diffusion coefficients of Li in Bi have been measured by Temnogorova et al.,
Weppner et al., van Norman and Newhouse, as shown in table \ref{t:diffusion} and
figure \ref{f:diffusivity}. Due to the available concentration
dependence, we use the formula of Newhouse. 

\begin{table}[h!]
\caption{Diffusivity of Li in Bi.}
\label{t:diffusion}
\centering
\small
\begin{tabular}{rrl}
\toprule
\multicolumn{1}{c}{$D$ in cm$^2$/s} & \multicolumn{1}{c}{$\vartheta$ in $^\circ$C} & \multicolumn{1}{c}{source}\\
\midrule
$2.2 \cdot 10^{-5}$ & 450 & \cite{vanNorman1961}\\
$1.56\pm 0.15 \cdot 10^{-5}$ & 500 & \cite{Temnogorova1979}\\ 
$3.25\pm 0.53 \cdot 10^{-5}$ & 550 & \cite{Temnogorova1979}\\
$3.27\pm 0.3 \cdot 10^{-5}$  & 550 & \cite{Temnogorova1979}\\
various (Li$_3$Bi) & 360-600 & \cite{Weppner1977a}\\
$\exp\left(\frac{-4.081 c - 0.01315}{c^2 + 0.3742 c +
  0.001572}\right)$ & 450 & \cite[p. 174]{Newhouse2014}\\
\bottomrule
\end{tabular}\end{table}

\subsection{Thermodynamic properties}
The thermodynamics of the Li-Bi system have been studied several times
-- table \ref{t:thermodynamics} gives an overview on literature sources
for the activity $a_\mathrm{Li(Bi)}$, the Gibbs free energy $G$, the equilibrium cell potential ($E_\mathrm{eq}$), the enthalpy $H$ and the entropy $S$. 
\begin{table}
\caption{Sources for thermodynamic properties of the Li-Bi system:
  activity ($a_\mathrm{Li(Bi)}$), Gibbs free energy ($G$), equilibrium cell potential ($E_\mathrm{eq}$),
  entropy of mixing ($S$) and enthalpy of mixing
  ($H$).}\label{t:thermodynamics}
\centering
\tiny
\begin{tabular}{ccccccl}
\toprule
\kz{phase diagram} &
\kz{$a_\mathrm{Li(Bi)}$} &
\kz{$G$} &
\kz{$E_\mathrm{eq}$} &
\kz{$S$} &
\kz{$H$} &
\kz{literature}\\
\midrule
$\bullet$ & $\bullet$ & $\bullet$ &  & $\bullet$ & $\bullet$ & \cite{Liu2013b}\\
 &  &  & $\bullet$ &  &  &   \cite[p. 152]{Vogel1964}\\
  &  & $\bullet$ &  &  &  &   \cite[p. 118]{Lawroski1963b}\\
  &  &  & $\bullet$ &  &  &   \cite[p. 216]{Lawroski1963}\\
$\bullet$ &  &  &  &  &  &   \cite{Ning2015}\\
  &  &  & $\bullet$ &  &  &   \cite[p. 226]{Lawroski1963a}\\
$\bullet$ &  & $\bullet$ & $\bullet$ &  &  &   \cite{Foster1964}\\
$\bullet$ &  & $\bullet$ & $\bullet$ & $\bullet$ & $\bullet$ &   \cite{Gasior1994}\\
  &  &  & $\bullet$ &  &  &   \cite{Saboungi1978}\\
 $\bullet$ & $\bullet$ & $\bullet$ &  & $\bullet$ & $\bullet$   & \cite{Cao2014}\\
  & $\bullet$ & $\bullet$ & $\bullet$ & $\bullet$ & $\bullet$   & \cite{Demidov1973}\\
  &  &  & $\bullet$ &  &  &   \cite[p. 109f]{Cairns1967}\\
  &  &  & $\bullet$  &  &  &   \cite[p. 143f]{Crouthamel1967}\\
 &  &  & $\bullet$ &  &  &   \cite[p. 110f]{Newhouse2014}\\
&&&$\bullet$&&&\cite{Weppner1978}\\
$\bullet$&$\bullet$&&&&$\bullet$&\cite{Predel1992}\\
&$\bullet$&&&&&\cite{vanNorman1961}\\
&$\bullet$&&$\bullet$&&&\cite{Wang2019}\\
\bottomrule
\end{tabular}\end{table}

\section{\new{Results}}
\subsection{Small Li-Bi cell}\label{s:smallCell}
\begin{figure}[h]
\centering
\includegraphics[width=0.9\textwidth]{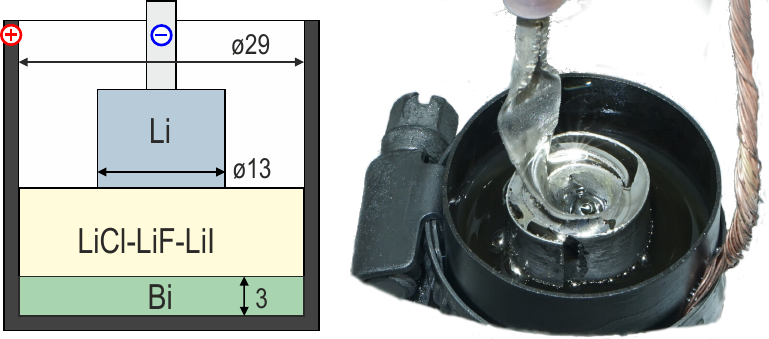}
\caption{Setup and dimensions (in mm) of the small Li-Bi laboratory
  cell (not to scale). The thicknesses of positive electrode and electrolyte are given for the
  start of the charge phase (figure \ref{f:08}).}\label{f:cell1Setup}.
\end{figure}

The first use case for the developed model is illustrated in
figure \ref{f:cell1Setup} -- a small Li-Bi cell with a diameter of
approximately 3\,cm. The Li-metal is contained in spirally rolled
Ni-sheet serving as current collector and separated from the positive electrode
by an effectively 3-4\,mm thick eutectic LiCl-LiF-LiI electrolyte.

The metals (Li: Alfa Aesar, 99.9\,\%; Bi: Alfa Aesar, 99.998\,\%) were
first cleaned mechanically (oil removal and skin cut off for Li),
before melting them and skimming floating pollutants. Additionally, a
small amount of salt was added to bind remaining impurities. After freezing
the contaminated salt, the clean metal was poured into a second
crucible. To prepare the electrolyte, LiCl (Alfa Aesar, 99.995\%) and
LiF (Beantown, 99.99\%) were mixed and allowed to rest for several
hours at 550\,$^\circ$C. LiI (Beantown, 99.95\%) was added thereafter,
and the complete mixture filtered though a glass frit.

The tantalum crucible was filled with 0.1\,mol Bi and 18.5\,g
electrolyte. Finally, molten Li was soaked into the negative electrode
current collectors (Ni alloy 200). The cell was operated on a ceramic
heating plate (BACH Resistor Ceramics) in a glove box filled with
argon gas (H$_2$O $<$ 0.1\,ppm, O$_2$ $<$ 0.1\,ppm) at a heating plate
temperature of $460^\circ$C \cite{Personnettaz2019}.

\begin{figure}[h]
\centering
\includegraphics[width=0.6\textwidth]{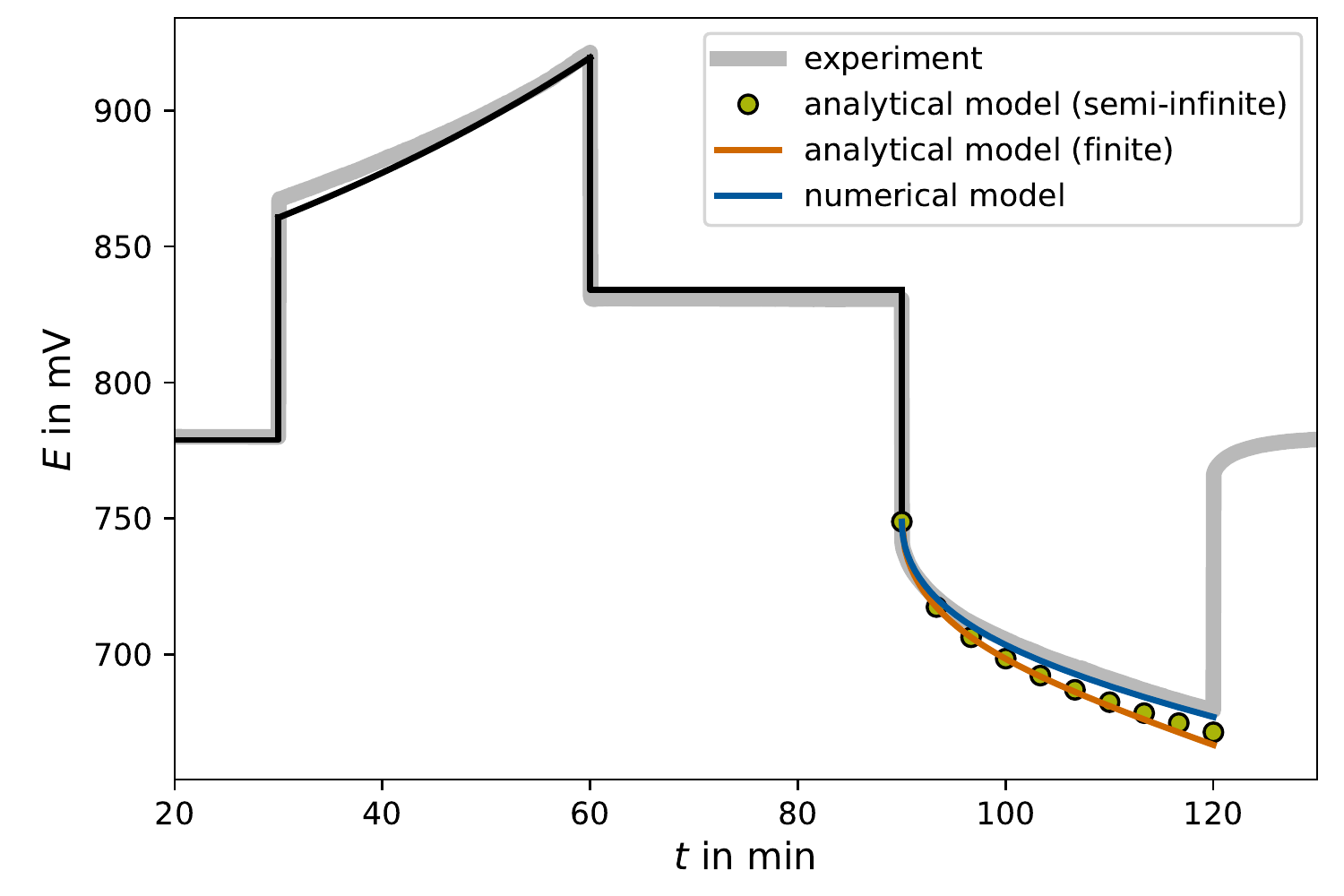}
\caption{Measured charge-discharge curve at 1\,A and modelled curves
  using the analytical solution for a finite and semi-infinite layer
  and the numerical   solution with variable diffusivity and volume
  change.}\label{f:08}
\end{figure}
The modelling work starts with the initial condition: the known mass
of Bi, and the measured equilibrium cell potential lead to the initial
Li-concentration in Bi. Then, the amount of Li, which leaves the
positive electrode during charge, is calculated. Considering the strong solutal
convection during charging, perfect mixing is assumed
\cite{Personnettaz2019}. The ohmic loss for charging is obtained for each
time-step using equation (\ref{eqn:ohmicCylinderFit}) and taking the change 
of the electrolyte layer height into account; as the latter changes with
time, it is derived from the known conductivity and geometry. 

Switching to discharging, the current direction is reversed and Li
transferred back to the positive electrode. For comparison, the diffusion
equation is solved once by the analytical solutions for a finite and
semi-infinite layer and once numerically. The latter accounts for
volume change of positive electrode and electrolyte as well as the variable
diffusivity, while the analytical solutions neglect volume change at
all and use a constant diffusion coefficient of
$4\cdot10^{-5}$\,cm$^2$/s.

Comparing the measured and modelled curves, it becomes apparent that
an approximate reproduction of experimental curves is easy. Already
Personnettaz et al.~\cite{Personnettaz2019} reproduced the same experiment 
using an estimated diffusion coefficient of $7\cdot10^{-5}$\,cm$^2$/s and a different formulation for the mixture density. Obtaining a perfect
match is, however, challenging. For example, the small deviation of
the equilibrium cell potential before charge/discharge is probably caused
by the fit function (equation \ref{eqn:emf}). As no measured electromotive force
values were available for 460\,$^\circ$C, the fit had to be slightly
extrapolated. The small mismatch of the cell potential during charge
is surely related to the ohmic loss. While the positive electrode layer changes
its height by 5\,\% during cycling, it is not easy to predict the exact
shape of the negative electrode. The Li might be soaked fully into the Ni-sheet or
form a large droplet below it. The shape of this droplet might vary
due to the Lorentz forces at charge/discharge and also due to the
transfer of Li-metal \cite{Benard2021}. Finally, the slight deviation
of the discharge curves might be related to the three-dimensionality of
the diffusion problem. While the model assumes only downward
diffusion, Li will in reality also diffuse sidewards in the positive electrode --
which then increases the cell voltage. Moreover, any very small flow
effects, caused, e.g. by Marangoni convection or electro-vortex flow,
might lead to a higher cell potential, too. Anyway, the numerical
solution is extremely close to the measured cell potential.

\subsection{Large Li-Bi cell}\label{s:largeCell}
\begin{figure}[h]
\centering
\includegraphics[width=0.9\textwidth]{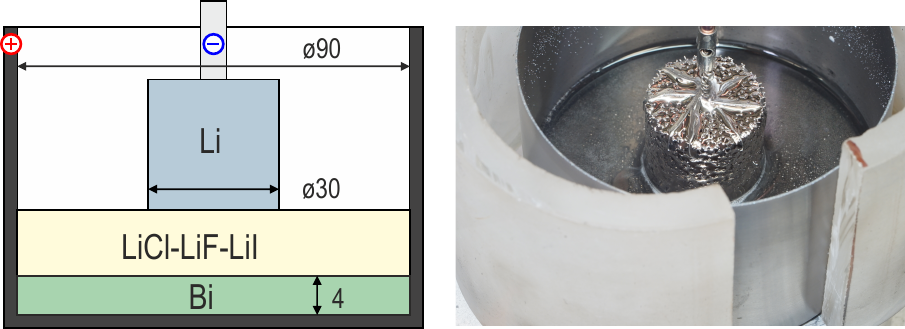}
\caption{Setup and dimensions (in mm) of the large Li-Bi cell (not to scale). The
  layer thicknesses are given for the start of discharge in figure \ref{f:12}.}\label{f:cell2Setup}
\end{figure}

Highlighting the broad applicability of the developed model, a second,
larger Li-Bi cell will be studied. Now, the cell diameter is 9\,cm,
with the negative electrode metal being contained in a Ni foam as shown in
figure \ref{f:cell2Setup}.

Metals and salts were cleaned as for the small cell (section
\ref{s:smallCell}). The battery was set up by pouring 1\,mol Bi into the
tantalum vessel and adding 80.28\,g eutectic LiCl-LiF-LiI
electrolyte. The negative current collector (Ni foam, Recemat BV) was prepared
by heating 1.5\,mol Li in a stainless steel vessel to 450\,$^\circ$C and
letting the foam rest for 2\,h in the bath. Finally, the cell was
operated such that the heating plate obtained a temperature of
500\,$^\circ$C, while the salt reached only 420\,$^\circ$C.

\begin{figure}[h]
\centering
\includegraphics[width=0.7\textwidth]{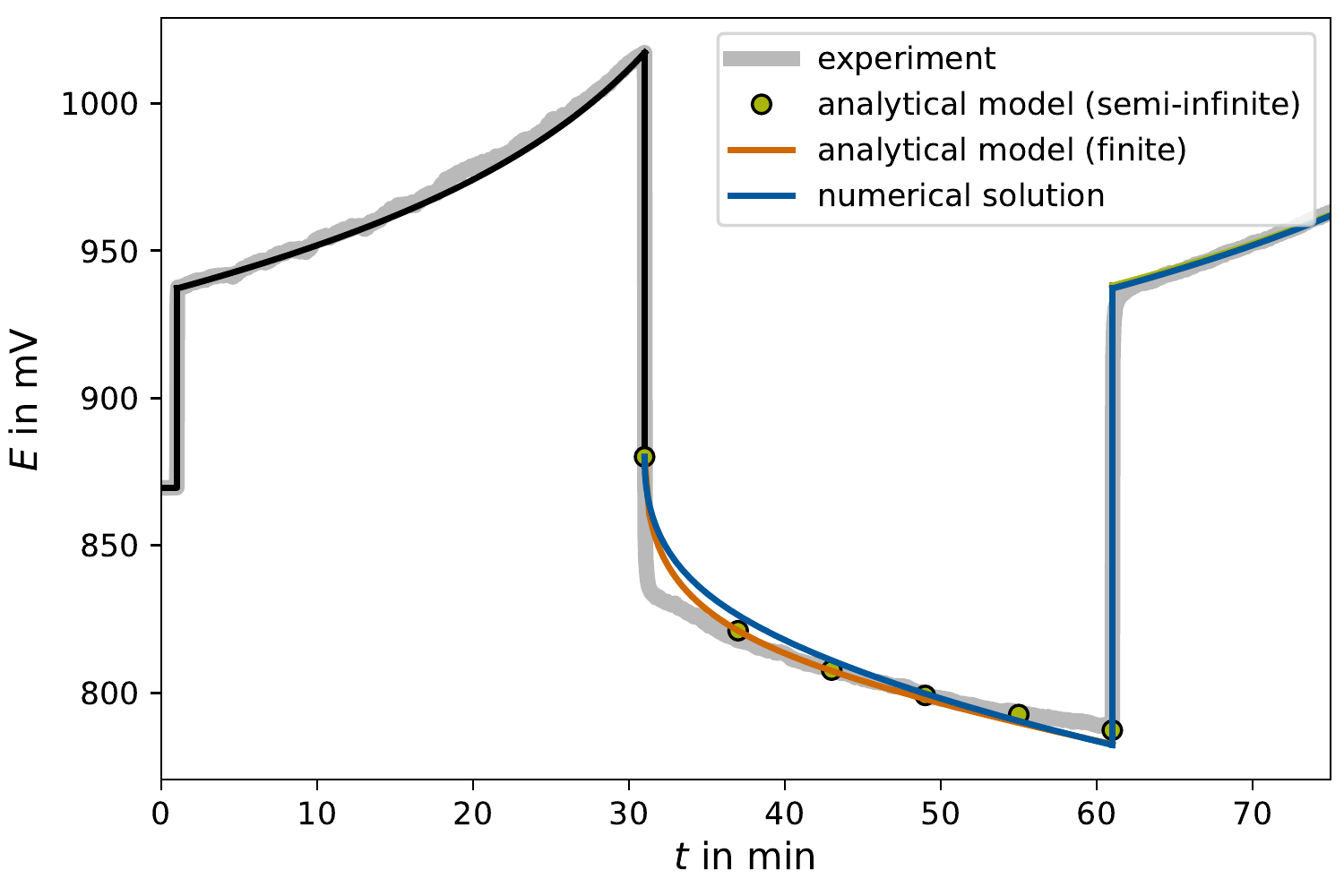}
\caption{Experimental and numerical charge-discharge curve for the
  large Li-Bi cell at 5\,A.}\label{f:12}
\end{figure}
The charge/discharge curve is modelled exactly like for the previous
experiment: the initial values are obtained from the measured
electromotive force and the positive electrode is assumed to be well mixed at
charge. Volume change is accounted for during charge, but during
discharge only when solving the diffusion equation numerically. As the
cell is operated with a Li-concentration of 1 to 5\,kmol/m$^3$, a
diffusion coefficient of $7\cdot10^{-5}$cm$^2$/s is assumed when using
the analytical solutions of the diffusion equation.

The model reproduces the experimental data very well -- with a
certain deviation during the discharge phase. Compared to the small
Li-Bi cell experiment described in section \ref{s:smallCell} the
charge phase is reproduced better. The reason is most likely that
the negative electrode shape is now defined much better, as the Li is contained in
a metal foam instead of a rolled Ni-sheet. The slight deviation during
cell discharge can most probably be attributed to the
three-dimensionality of the problem. As the positive electrode diameter is three
times larger than the negative electrode, substantial lateral diffusion will occur,
which surely influences the cell voltage. Nevertheless, the
experimental curve is reproduced reasonably well. 

\subsection{Li-Bi cell by Ning et al.}\label{s:Ning}
\begin{figure}[h]
\centering
\includegraphics[width=0.45\textwidth]{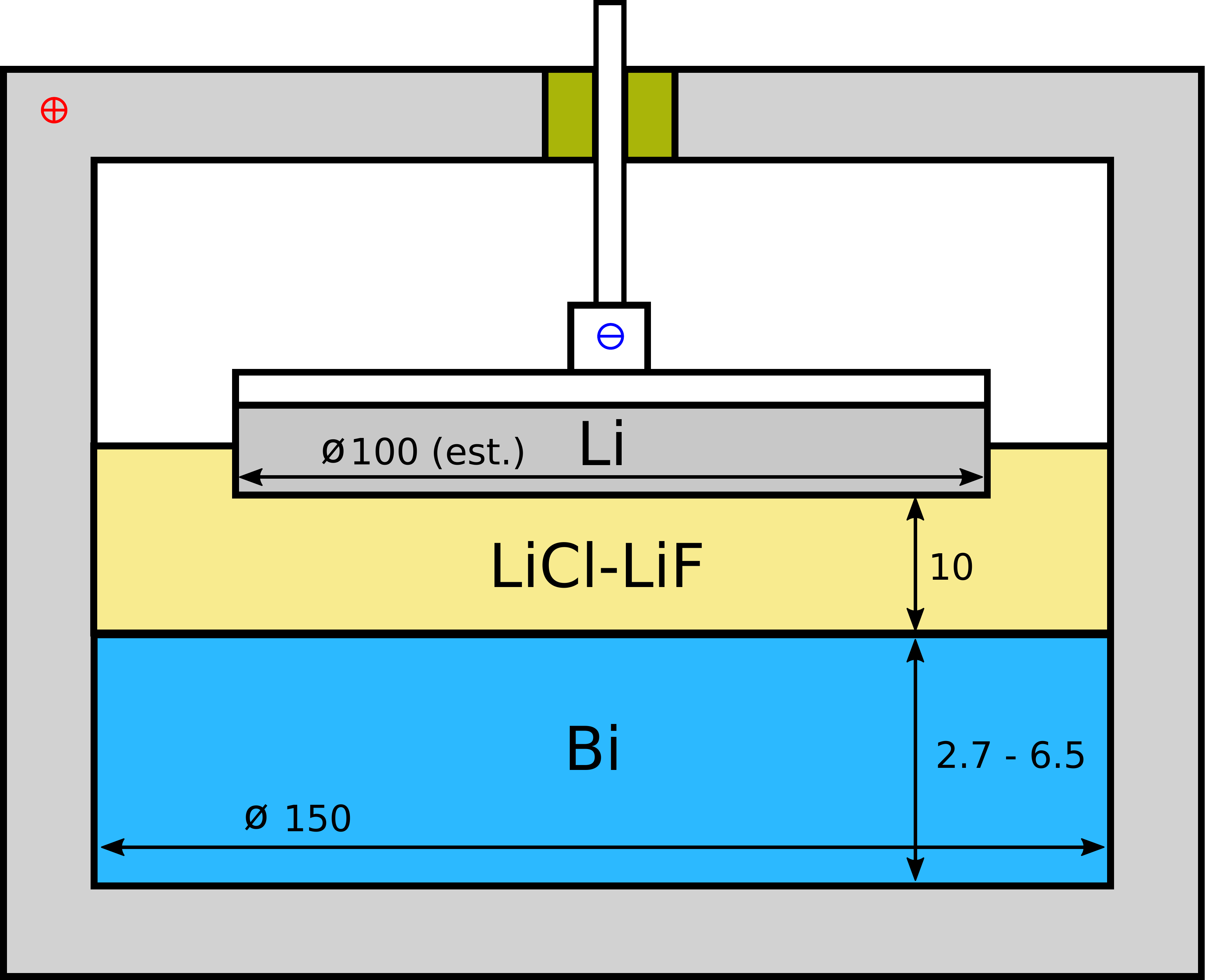}
\caption{Setup and dimensions (in mm) of the Li-Bi cell investigated
  by Ning et al. \cite{Ning2015} (not to scale).}\label{f:cell3Setup}
\end{figure}

The last test case is a large Li-Bi cell built at Massachusetts
Institute of Technology and published by Ning et
al. \cite{Ning2015}. The setup is shown in figure \ref{f:cell3Setup}:
the cell is 15\,cm in diameter, containing 455\,g of Bi, which results
in a 2.7\,mm thick positive electrode layer. The Li-negative electrode is soaked into a Ni
foam, and is assumed to be 10\,cm in diameter. The cell operates at
550\,$^\circ$C using an eutectic LiCl-LiF (70:30\,mol\%) electrolyte
with an interelectrode distance of 10\,mm in charged state. Having a
capacity of 175\,Ah, 3\,h and 20\,min are theoretically needed to
discharge the complete cell at a current density of 300\,mA/cm$^2$. 

\begin{figure}[h]
\centering
\includegraphics[width=0.8\textwidth]{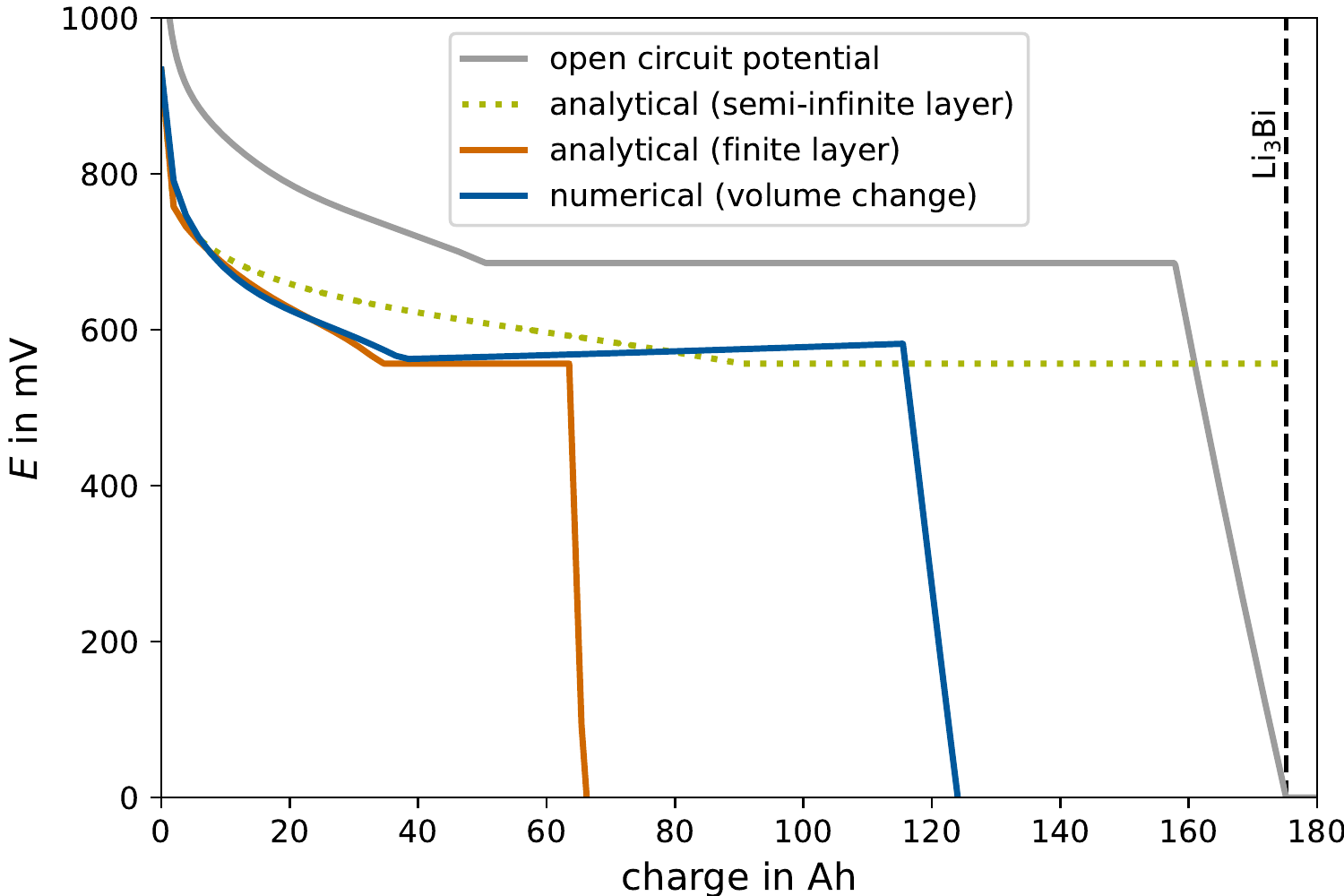}
\caption{Discharge curves for 0.3\,A/cm$^2$ using different numerical
  and analytical models.}\label{f:cell3Discharge}
\end{figure}

Figure \ref{f:cell3Discharge} illustrates the discharge curves of this
cell using different models. The equilibrium cell potential
(equation \ref{eqn:emfComplete}) -- which would be obtained at very small
current densities -- clearly exhibits a long plateau in the two-phase
area. When reaching a Li molar fraction of about 0.73, there is a
predicted extended intermetallic phase region \cite{Pavlyuk2015}. In
our model, the potential in this region decreases linearly to zero.

Discharging with a current density of 300\,mA/cm$^2$, the numerical
solution of the diffusion equation gives the most accurate result as
it accounts for volume change as well as the concentration dependent
diffusivity. As expected, the ohmic overpotential shifts the cell
potential to lower values. Moreover, the usable cell capacity is
reduced by roughly 30\,\%, as well. The reason is simply that due to
the high current, a layer of saturated Li$_3$Bi forms at the
electrolyte-positive electrode interface, which causes the cell potential to drop
to 0\,V. The ``missing'' capacity is located just below of this
intermetallic layer, where a stoichiometric composition of Li$_3$Bi is
not reached. The usable cell capacity depends therefore not only on
the mass of the active materials and side reactions, but also on the
discharge current.

Modelling the discharge curve with the analytical solution of the
diffusion equation for a finite layer (equation \ref{eqn:diffFinite}) using
an average diffusivity of $D=8\cdot 10^{-5}$\,cm$^2$/s
\cite{Weber2020}, the predicted capacity is only 50\% of that found
by the numerical solution. This considerable discrepancy is simply
caused by the fact that volume changes are not considered. As the analytical solution of
the diffusion equation neglects the fact that the positive electrode thickness
changes approximately by a factor of two (see
figure \ref{f:volumeChange}) between charge and discharge, it predicts a
much lower capacity.

The analytical solution of the diffusion equation for a semi-infinite
layer does the opposite: it predicts a much larger cell capacity than
theoretically possible. As it ignores the lower boundary of the
positive electrode, it assumes that much more Li can be alloyed into Bi than
practically possible. 

These examples highlight that the analytical
solutions for the diffusion equation should not be used when
considering full charge-discharge cycles as they induce considerable
errors due to the negligence of volume-change effects.

\section{Application scenarios and limitations of the model}
\subsection{Use cases and application scenarios}\label{s:useCases}
\new{The developed model can easily be applied to other liquid metal
batteries -- basically only the material properties need to be
changed. Possible use cases or application scenarios span from basic
to applied research and include, amongst others, the following:
\begin{itemize}
\item Preparation of experiments: battery cycling experiments can be
  presimulated, which helps to choose, e.g. an appropriate cycling
  time and cell current.
\item Reference and validation: the model can be used as reference
case or for validation purposes, e.g. when performing costly
three-dimensional simulations of complete batteries.
\item Cost assessment: the cell voltage is one of the most influential
  parameters of a battery cost model for LMBs \cite{Berridge2021}. The
  developed cell voltage model can make these estimations much more
  accurate.
\item Scale up and system modelling: the design of large battery
  stacks including hundreds of single cells for special use cases can
  largely be facilitated using optimisation algorithms, such as
  genetic algorithms. The developed cell voltage model could be
  integrated to provide the power, heat release, current and cell
  potential of the single cells. Especially, the simplicity and speed
  of our model would be very beneficial for such optimisation tasks.
\item Battery management systems: the model can be used as the core of
  a battery management system. By only measuring the temperature and
  open circuit potential, it might provide for example the state of
  charge, estimate the charge or discharge time, the remaining
  capacity or the model might be used to predict the cell potential
  as function of time and battery current.
\end{itemize}}

\subsection{Limitations and possible improvements}
\begin{figure}[h]
\centering
\subfigure[]{\includegraphics[height=0.35\textwidth]{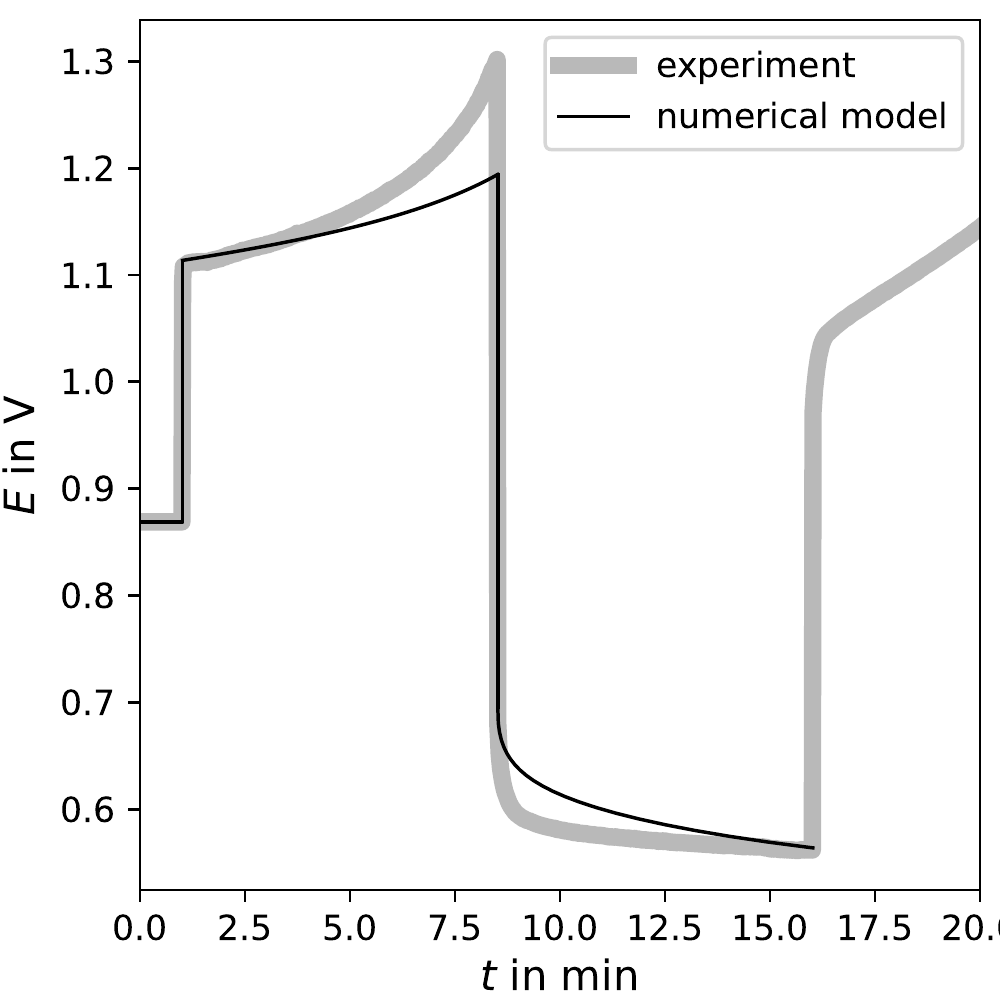}}
\subfigure[]{\includegraphics[height=0.35\textwidth]{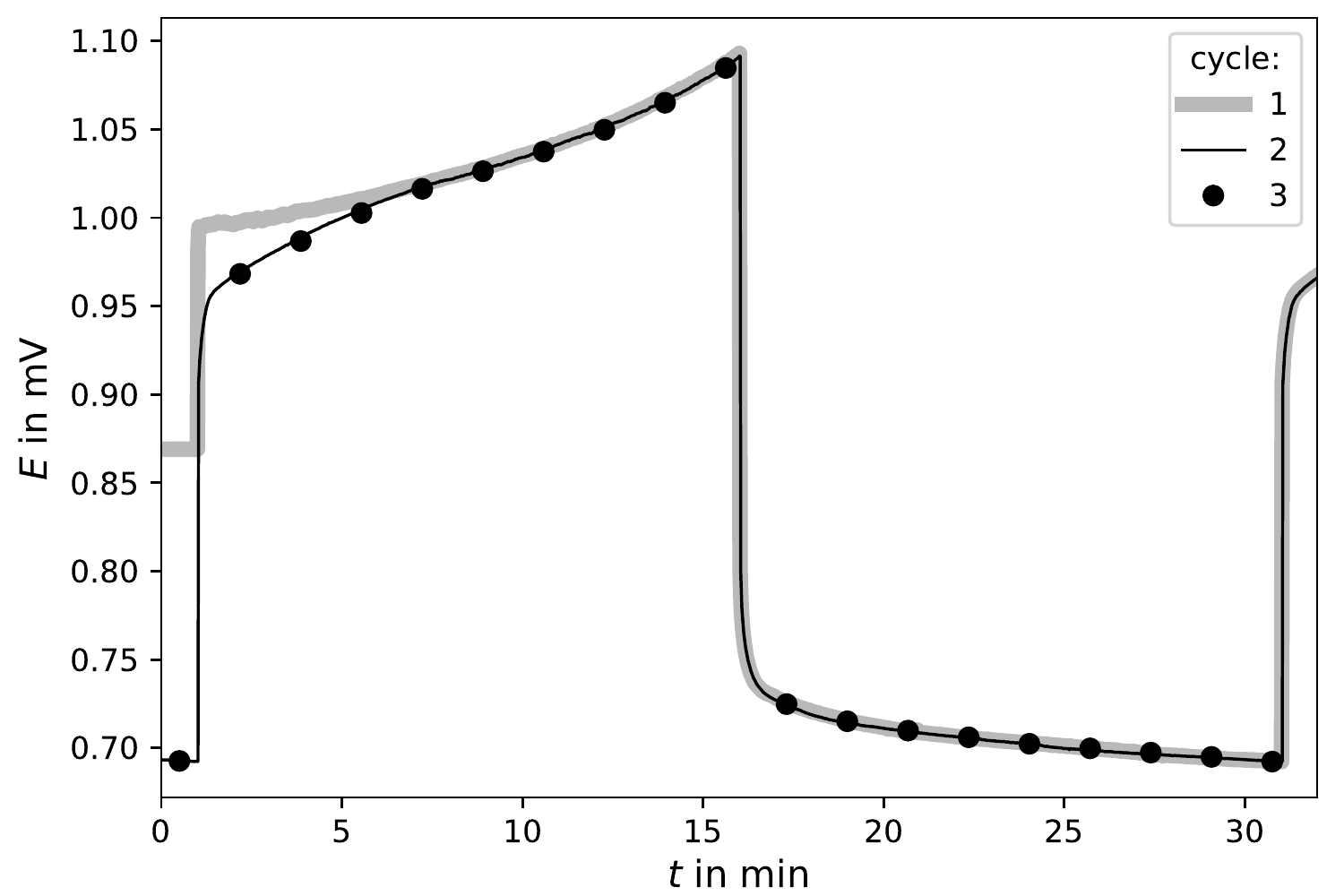}}
\caption{Experimental and modelled charge/discharge curve for the
  large Li-Bi cell presented in section \ref{s:largeCell} at 20\,A (a)
  and overlay of the first three experimental cycles at 10\,A (b).}\label{f:limitations}
\end{figure}

Modelling the cell voltage of LMBs is a complex task\new{, involving
numerous simplifications and uncertainties. However, most of them 
are negligible and do not restrict the field of application of
our model.} In the opinion
of the authors, the presented model and its portability to other
geometries and cell chemistries is mainly limited by the following
points:
\begin{enumerate}%#[label=(\arabic*)]
\item Material properties: measured densities of typical LMB electrode couples
  are usually scarce. Moreover, diffusivities are not always available
  with their concentration dependence, and the measurement error may
  be considerable.
\item Complexity of intermetallic phases: the conductivity of the Li-Bi
  alloy changes by almost two orders of magnitude during alloying,
  reaching the same order of magnitude as for the molten salt, when
  forming the intermetallic phase \cite{Steinleitner1975,Nguyen1977,Grube1934}. The
  corresponding ohmic losses might be included in improved
  models. Likewise, the diffusivity of Li in Bi changes by a factor of
  up to 3, when forming the intermetallic. This makes accurate
  predictions of mass transport challenging.
\item Ohmic overpotential: the ohmic losses in the salt can usually
  not be approximated by a one-dimensional model. The fit equation, used
  here, might be replaced alternatively by solving a Laplace equation
  for the electric potential in 2D or 3D.
\item One-dimensionality of mass transport: mass transport has been
  modelled as a one-dimensional effect. This simplification is often
  appropriate -- as long as the negative electrode and positive electrode diameters are
  similar (see section \ref{s:largeCell} for an example, where this
  was not   the case).
\item Simplification of flow effects: any type of flow can increase mass transfer
  \cite{Weber2020}. This effect
  has been neglected. The positive electrode is simply be assumed to be
  perfectly mixed by solutal convection at charge
  \cite{Personnettaz2019}, while mass transfer is assumed to be
  controlled by diffusion only at discharge. One example, where this
  does not work well, is shown in figure \ref{f:limitations}(a). When
  charging the large Li-Bi cell described in section \ref{s:largeCell}
  with a very large current (here 20\,A), a Li-depleted concentration
  layer will form in the positive electrode despite most of the volume
  is well mixed by solutal convection. The measured cell potential
  will rise therefore steeply, while the modelled potential is lower.
\item Memory effect: the cell potential is not only a function of the
  state of charge, but is history-dependent as well.
  Figure \ref{f:limitations}(b) shows one example: the measured cell
  potential of the first cycle is larger than the one of the two subsequent
  ones. At the beginning of cycling, the positive electrode was at rest and perfectly
  mixed. Therefore, solutal convection sets in immediately at charge,
  mixing the alloy well, and leading to a high cell voltage. After
  discharge, a stable density stratification formed. When now charging
  again, solutal convection takes longer to destroy the stable density
  stratification and to mix the positive electrode again -- therefore, the potential
  of the second and third cycle is initially lower. Such ``memory
  effects'' can generally be accounted for using the developed model;
  however, care needs to be taken not to ignore them.
\end{enumerate}

\section{Summary}
The Lithium-Bismuth system is one of the best explored liquid metal
batteries (LMBs). After discussing the peculiarities of this chemistry
and summarising the previous experimental studies, we gave a short
overview of the most relevant modelling work of LMBs. Further, we
briefly described our motivation to develop a simple \new{quasi-}one-dimensional
cell voltage model. 

The equilibrium cell potential is obtained by a complex two-dimensional
fit of experimental data as function of temperature and Li-molar
fraction. While the very small activation losses have been neglected,
different sub-models for the ohmic loss have been proposed. It was
shown that the voltage drop in the salt can be solved analytically,
but only under certain assumptions. As alternative, two fit functions
for cylindrical and rectangular cells are provided, which describe the
ohmic losses well.

Special care was taken to obtain the correct mass transfer
overpotential. After discussing the possible formulations of the
diffusion equation to obtain the Li-content in Bi, different
possibilities to convert molar and mass fraction to molar or mass
concentration were proposed. Two analytical solutions for the
diffusion equation were presented: one for a semi-infinite, and one
for a finite positive electrode. Finally, a numerical scheme was implemented in the finite
difference and finite volume method, which accounts for
volume change of the positive electrode during operation.

The developed model was finally applied to three different Li$||$Bi
cells. Although the results generally match well with the experimental
data, the comparison revealed certain limitations of the model. Most
importantly, the one-dimensional model fails in describing
three-dimensional mass transfer effects, which become important if the
negative electrode and positive electrode diameter deviate strongly from each other. Moreover,
the influence of flow effects on the cell voltage has been simplified
to a large extend. The modelling of the three experiments further highlighted that
for short-time cycling the analytical solutions of the diffusion
equation are well suited. However, in order to describe the full
discharge of an LMB, volume change is of highest importance, and needs
to be included as well by solving the mass transport equation
numerically.

\section*{Software}
The raw data, python model and examples can be obtained from
\url{https://doi.org/10.14278/rodare.1369}.

\section*{Acknowledgement}
This project has received funding from the European Union’s Horizon
2020 research and innovation programme under grant agreement No
963599, from the Deutsche Forschungsgemeinschaft (DFG, German
Research Foundation)  by  award  number  338560565 and in frame of the
Helmholtz - RSF Joint Research Group ``Magnetohydrodynamic
instabilities: Crucial relevance for large scale liquid metal
batteries and the sun-climate connection'', contract No.~HRSF-0044 and
RSF-18-41-06201. We would like to thank A. Linke and J. Fuhrmann for
fruitful discussions about the solution of the diffusion equation.

\bibliography{literature}

\appendix
\section{Ohmic overpotential}
\subsection{Analytical solution for cylindrical geometries}\label{a:ohmAnalytical}
We calculate the potential distribution in the electrolyte for the
left axisymmetric cylindrical cell shown in Figure \ref{f:foam}. \new{In
  reality, even the sidewalls of the negative electrode will be
  covered to a certain extend by molten salt. As the current path from
  these side walls to the positive electrode is much longer than from
  the bottom of the negative electrode, we assume that all current
  flows through the bottom of the metal foam only.} 
Since
the electrical current \new{in the electrolyte} can be uniquely described by a potential
through Ohm's law $\bm{j} = -\sigma \bm{\nabla}\phi$ and must be closed
($\bm{\nabla}\cdot \bm{j}$), we need to solve the two-dimensional
Laplace equation 
\begin{equation}
        \Delta \phi = \frac{\partial^2 \phi}{\partial r^2} + \frac{1}{r}\frac{\partial \phi}{\partial r} + \frac{\partial^2 \phi}{\partial z^2} = 0\label{eq:Lap}
\end{equation}
in the electrolyte's domain, where we have applied cylindrical
coordinates ($r,\varphi,z$) with the origin located in the centre of
the salt-negative electrode interface. For the boundary conditions, we
assume that the sidewalls \new{between electrolyte and cell housing} are perfectly
insulating and that we have a homogeneous vertical current
distribution in the positive electrode \new{-- for a comparison of
different alternative boundary conditions, such as the Millere profile, see \cite{Millere1980,Bojarevics1989,Herreman2019b}.}
The boundary conditions read
\begin{align}
        &\left.\frac{\partial \phi}{\partial r}\right|_{r = R_1} = 0, \label{eq:Con1} \\
        &\left.\frac{\partial \phi}{\partial z}\right|_{z = -H} = \frac{I}{\sigma \pi R_1^2}, \label{eq:Con2} \\
        &\left.\frac{\partial \phi}{\partial z}\right|_{z = 0} = \frac{I}{\sigma \pi R_2^2}\Theta(R_2-r), \label{eq:Con3}
\end{align}
where $I$ denotes the total cell current and $\Theta$ is the Heaviside
function ensuring that no current bypasses the negative electrode in
the region $r > R_2$. Please note that the current distribution is
usually not homogeneous, in particular not in the negative electrode foam if its radius $R_2$ is considerably smaller than the cylinder radius $R_1$. However, we seek for an approximate solution of the cell voltage to be expressed analytically, requiring to keep the boundary conditions as simple as possible. This approximation is \textit{a posteriori} justified in figure \ref{f:ohmicLoss}, where we discuss the relative error following from this approximation.
The cylindrical boundary value problem (\ref{eq:Lap}) - (\ref{eq:Con3}) is solved by expanding a Fourier-Bessel series. We take the ansatz
\begin{equation}
        \phi = -\frac{I}{\sigma \pi R_1^2}z + \sum_{n=1}^{\infty}a_n \cosh\left(\frac{\epsilon_{0n}}{R_1}(z + H)\right)J_0 \left(\epsilon_{0n}\frac{r}{R_1}\right), \label{eq:Ansatz}
\end{equation}
with $J_0$ being the zero-order Bessel function of the first kind and the numbers $\epsilon_{0n}$ given as the $n$ roots of the first derivative of the zero-order Bessel function
\begin{equation}
        J_0^{'}(\epsilon_{0n}) = 0,
\end{equation}
which can be easily determined
numerically or found in, e.g. \cite{Abramowitz2013}. The coefficients $a_n$ are to be determined by applying condition (\ref{eq:Con3}); ansatz (\ref{eq:Ansatz}) already fulfils the simple boundary conditions (\ref{eq:Con1}) and (\ref{eq:Con2}). Inserting ansatz (\ref{eq:Ansatz}) into (\ref{eq:Con3}), multiplying both sides of the equation by $rJ_0(\epsilon_{0m}r/R_1)$ and eventually integrating the equation over the cylinder radius $R_1$, we find explicit expressions for the coefficient $a_n$ as
\begin{equation}
        a_n = \frac{2IJ_1\left(\epsilon_{0n} \frac{R_2}{R_1}\right)}{\sigma \pi R_2 \sinh\left(\frac{\epsilon_{0n}}{R_1}H\right)\epsilon_{0n}^2 J_0^2(\epsilon_{0n})}. \label{eq:Coeff}
\end{equation}
For the calculation, we have exploited the orthogonality condition
\begin{equation}
        \int_0^{R_1} r J_0\left(\epsilon_{0m} \frac{r}{R_1}\right)J_0\left(\epsilon_{0n} \frac{r}{R_1}\right) {\rm d}r = \frac{1}{2} R_1^2 J_0^2(\epsilon_{0n})\delta_{mn} .
\end{equation}
Inserting (\ref{eq:Coeff}) into (\ref{eq:Ansatz}) finally yields the potential solution
\begin{align}
        \phi = -\frac{I}{\sigma \pi R_1^2}z - \sum_{n=1}^{\infty}\frac{2IJ_1\left(\epsilon_{0n} \frac{R_2}{R_1}\right)}{\sigma \pi \epsilon_{0n}^2 R_2}\frac{ \cosh\left(\frac{\epsilon_{0n}}{R_1}(z + H)\right)}{\sinh\left(\frac{\epsilon_{0n}}{R_1}H\right)} \frac{J_0\left(\epsilon_{0n}\frac{r}{R_1}\right)}{J_0^2\left(\epsilon_{0n}\right)}.
\end{align}
The solution contains the voltage loss expected to occur in the
electrolyte. It is defined as the potential difference between
negative and positive electrode, but we must take into account that
the potential is not constant along the negative electrode foam. To approximate the
global cell voltage, we consider the mean potential in the negative electrode (the positive electrode potential is approximately constant), so that the voltage drop is given as
\begin{align}
        \eta_\Omega &:= \frac{1}{R_2}\int_{0}^{R_2}\phi(r,z=-H) - \phi (r,z=0){\rm d}r \nonumber \\
        &= \frac{IH}{\sigma \pi R_1^2} + \sum_{n=1}^{\infty}\frac{IJ_1\left(\kappa_{0n}\right)}{\sigma \pi \epsilon_{0n}^2 R_2}\tanh\left(\frac{\epsilon_{0n}}{R_1}\frac{H}{2}\right) \frac{\pi H_0\left(\kappa_{0n}\right)J_1\left(\kappa_{0n}\right) + \left(2 - \pi H_1\left(\kappa_{0n}\right) \right)J_0\left(\kappa_{0n}\right)}{J_0^2\left(\epsilon_{0n}\right)}, \label{eq:Pot}
\end{align}
with $\kappa_{0n} = \epsilon_{0n}R_2 /R_1$. For the calculation we
have applied the hyperbolic identity $\coth(x) - \sinh(x)^{-1} =
\tanh(x/2)$. $H_0$ and $H_1$ refer here to the Struve functions of
zero and first order, see \citep{Abramowitz2013}, which result from
the integration of $J_0$. The first term in (\ref{eq:Pot}) describes
the homogeneous voltage for the trivial case $R_1 = R_2$. The
following Fourier sums correct this solution to account for the
inhomogeneities caused by the asymmetry of positive and negative
electrode $R_2 < R_1$. The convergence behaviour of the series
strongly depends on the ratio $R_2 / R_1$. For point-like negative electrodes $R_2 \ll R_1$ the solution converges very slowly but correspondingly fast if the electrodes are in the same order. For all the practical cases shown in figure \ref{f:ohmicLoss} it was sufficient to keep the first 400 terms allowing for a very fast calculation.

\subsection{Numerical solution for cylindrical geometries}\label{a:ohmicCylinderFit}
The voltage loss in the electrolyte of a cylindrical cell, as
illustrated in figure \ref{f:foam} is
\begin{equation}\label{eqn:ohmicCylinderFit}
\begin{split}
\eta_\Omega = I\frac{0.05^2}{\sigma R_1^2}(a &+ a_1A + a_2A^2 + a_3A^3 +
  a_4A^4 + a_5A^5 + a_6A^6\\
 &+ b_1H     + b_2H^2 + b_3H^3 + b_4H^4 + b_5H^5 + b_6H^6\\
 &+ d_1AH + d_2(AH)^2 +     d_3(AH)^3 + d_4(AH)^4 +
  d_5(AH)^5\\
  &+ e_1H/A + e_2H^2/A + e_3HA^2 + e_4AH^2 + e_5AH^3),
\end{split}
\end{equation}
with the ratio of the radii defined as $A=R_2/R_1$ and
\begin{equation}
\begin{split}
&a=12.5,\ a_1=-105,\ a_2=358,\ a_3=-652\ a_4=676,\ a_5=-378,\ a_6=89,\\
&b_1=214,\
b_2=-5.8\cdot 10^4,\
b_3=3.3\cdot 10^6,\
b_4=-8\cdot 10^7,\
b_5=1.3\cdot 10^9,\
b_6=-2.8\cdot 10^9,\\
&d_1=-1277,\
d_2=-4507,\
d_3=-4.3\cdot 10^6,\
d_4=2.4\cdot 10^8,\
d_5=-4.5\cdot 10^9,\\
&e_1=419,\
e_2=-1.1\cdot 10^4,\
e_3=691,\
e_4=9.6\cdot 10^4,\
e_5=-2\cdot 10^7.
\end{split}
\end{equation}
The formula is valid for $A> 0.3$ and electrolyte layers thicker than 2.5\,mm.

\subsection{Numerical solution for rectangular geometries}\label{a:ohmicSquareFit}
The voltage loss in the electrolyte of a square cell, as
illustrated in figure \ref{f:foam} is
\begin{equation}
\begin{split}
\eta_\Omega = I\frac{0.05^2}{\sigma L_1^2}(a &+ a_1A + a_2A^2 + a_3A^3 +
  a_4A^4 + a_5A^5 + a_6A^6\\
 &+ b_1H     + b_2H^2 + b_3H^3 + b_4H^4 + b_5H^5 + b_6H^6\\
 &+ d_1AH + d_2(AH)^2 +     d_3(AH)^3 + d_4(AH)^4 +
  d_5(AH)^5\\
  &+ e_1H/A + e_2H^2/A + e_3HA^2 + e_4AH^2 + e_5AH^3),
\end{split}
\end{equation}
with the ratio of the side lengths defined as $A=L_2/L_1$ and
\begin{equation}
\begin{split}
&a=45.5,\ a_1=-383,\ a_2=1312,\ a_3=-2395\ a_4=2485,\ a_5=-1390,\ a_6=326,\\
&b_1=354,\
b_2=-1.6\cdot 10^5,\
b_3=1\cdot 10^7,\
b_4=-2.6\cdot 10^8,\
b_5=4.2\cdot 10^9,\
b_6=-1.3\cdot 10^{10},\\
&d_1=-3384,\
d_2=3.9\cdot 10^4,\
d_3=-1.4\cdot 10^7,\
d_4=7.5\cdot 10^8,\
d_5=-1.3\cdot 10^{10},\\
&e_1=1393,\
e_2=-3.5\cdot 10^4,\
e_3=1746,\
e_4=2.3\cdot 10^5,\
e_5=-5.1\cdot 10^6.
\end{split}
\end{equation}
The formula is valid for $A>0.2$ and electrolyte layers, which are at least 2\,mm thick.

\section{Discretisation of the diffusion equation}
\subsection{Finite difference method}\label{a:diffusionNumericalFDM}
\subsubsection{Equation}
The diffusion equation
\begin{equation}
\frac{\partial c}{\partial t} = \nabla\cdot D\nabla c
\end{equation}
is discretised using central differencing and the implicit Euler method as \cite{Ferziger2008} (see also \cite{Langtangen2017})
\begin{equation}
\frac{c^t - c^{t-1}}{\Delta t} = \frac{\left(D\frac{\partial c}{\partial z}\right)_{i+0.5} - \left(D\frac{\partial c}{\partial z}\right)_{i-0.5}}{0.5(z_{i+1} - z_{i-1})}
\end{equation}
with $t$ denoting the time, $i$ the point-index and $z$ the coordinate
running from bottom to top. The first derivatives are discretised as
\begin{equation}
\left(D\frac{\partial c}{\partial z}\right)_{i+0.5} = \frac{D_i + D_{i+1}}{2}\cdot \frac{c_{i+1}-c_i}{z_{i+1} - z_i},
\end{equation}
and
\begin{equation}
\left(D\frac{\partial c}{\partial z}\right)_{i-0.5} = \frac{D_{i-1} + D_i}{2}\cdot \frac{c_i-c_{i-1}}{z_i - z_{i-1}}.
\end{equation}
Combining these equations leads to
\begin{equation}
c^t = \frac{\frac{D_i + D_{i+1}}{2}\cdot \frac{c_{i+1}-c_i}{z_{i+1} - z_i} - \frac{D_{i-1} + D_i}{2}\cdot \frac{c_i-c_{i-1}}{z_i - z_{i-1}}}{0.5(z_{i+1} - z_{i-1})}\Delta t + c^{t-1}
\end{equation}
and after further simplification to
\begin{equation}
c^t = \frac{\frac{D_i + D_{i+1}}{2}\cdot (c_{i+1}-c_i)\cdot(z_i - z_{i-1}) - \frac{D_{i-1} + D_i}{2}\cdot (c_i-c_{i-1})\cdot(z_{i+1} - z_i)}{0.5(z_{i+1} - z_{i-1})\cdot(z_{i+1} - z_i)\cdot(z_i - z_{i-1})}\Delta t + c^{t-1}
\end{equation}
and finally to
\begin{align}
\begin{split}
c^t &= c_{i-1}\cdot \frac{\frac{D_{i-1} + D_i}{2}\cdot (z_{i+1} - z_i)}{0.5(z_{i+1} - z_{i-1})\cdot(z_{i+1} - z_i)\cdot(z_i - z_{i-1})}\Delta t\\[0.2em]
    &+ c_i\cdot\frac{-\frac{D_i + D_{i+1}}{2}\cdot(z_i - z_{i-1}) - \frac{D_{i-1} + D_i}{2} \cdot(z_{i+1} - z_i)}{0.5(z_{i+1} - z_{i-1})\cdot(z_{i+1} - z_i)\cdot(z_i - z_{i-1})}\Delta t\\[0.2em]
    &+ c_{i+1}\cdot\frac{\frac{D_i + D_{i+1}}{2} \cdot(z_i - z_{i-1}) }{0.5(z_{i+1} - z_{i-1})\cdot(z_{i+1} - z_i)\cdot(z_i - z_{i-1})}\Delta t\\[0.2em]
    &+ c^{t-1}.
\end{split}
\end{align}

The equation is simplified by setting $c^t = c_i$, defining
\begin{equation}
a = \frac{\Delta t}{(x_{i+1} - x_{i-1})\cdot(x_{i+1} - x_i)\cdot(x_i - x_{i-1})}
\end{equation}
and rearranging to fit to the matrix equation
\begin{equation}
A\cdot c = b,
\end{equation}
which leads to
\begin{align}
\begin{split}
&c_{i-1}\cdot -(D_{i-1} + D_i)\cdot (z_{i+1} - z_i)\cdot a\\[0.2em]
&+ c_i\left(1 + \cdot((D_i + D_{i+1})\cdot(z_i - z_{i-1}) + (D_{i-1} + D_i) \cdot(z_{i+1} - z_i))\cdot a\right)\\[0.2em]
&+ c_{i+1}\cdot - (D_i + D_{i+1}) \cdot(z_i - z_{i-1})\cdot a\\[0.2em]
&= c^{t-1}.
\end{split}
\end{align}
We now define the upper distance
\begin{equation}
dz_u = z_{i+1} - z_i,
\end{equation}
and lower distance
\begin{equation}
dz_l = z_i - z_{i-1}
\end{equation}
and
\begin{equation}
dz_u + dz_l = z_{i+1} - z_{i-1}.
\end{equation}
This gives
\begin{equation}
a = \frac{\Delta t}{(dz_u + dz_l)dz_udz_l},
\end{equation}
which leads to
\begin{align}
\begin{split}
&c_{i-1}\cdot -(D_{i-1} + D_i)\cdot dx_u\cdot a\\[0.2em]
&+ c_i\cdot\left(1 + ((D_i + D_{i+1})\cdot dx_l + (D_{i-1} + D_i) \cdot dx_u)\cdot a\right)\\[0.2em]
&+ c_{i+1}\cdot - (D_i + D_{i+1}) \cdot dx_l \cdot a\\[0.2em]
&= c^{t-1}.
\end{split}
\end{align}

\subsubsection{Boundary conditions}
The boundary condition at the lower interface reads
\begin{equation}
\nabla c\cdot\bi n = 0 = -c_1 + c_0
\end{equation}
which leads to the coefficients $A[0,0] = 1$, $A[0,1] = -1$ and $b=0$.

The boundary condition at the upper interface reads
\begin{equation}
\nabla c\cdot\bi n =\frac{j}{\nu_eFD}.
\end{equation}
This Neumann boundary condition is discretised as \cite{Ferziger2008}
\begin{equation}
\nabla c= \frac{-c_3(z_2-z_1)^2 + c_2(z_3 - z_1)^2 -c_1((z_3-z_1)^2 - (z_2-z_1)^2)}{(z_2-z_1)(z_3-z_1)(z_3-z_2)}.
\end{equation}
As $i$ counts from the bottom, we denote the last point by $-1$, the
second last point by $-2$ and so on, and find
\begin{equation}
\nabla c= \frac{-c_{-3}(z_{-2}-z_{-1})^2 + c_{-2}(z_{-3} - z_{-1})^2 -c_{-1}((z_{-3}-z_{-1})^2 - (z_{-2}-z_{-1})^2)}{(z_{-2}-z_{-1})(z_{-3}-z_{-1})(z_{-3}-z_{-2})}.
\end{equation}
We obtain the matrix coefficients as
\begin{equation}
b = \frac{j}{\nu_eFD}
\end{equation}
and
\begin{align}
A[-1, -3] &= -(z_{-2}-z_{-1})^2 / a \\
A[-1, -2] &=  (z_{-3} - z_{-1})^2 / a \\
A[-1, -1] &= -((z_{-3}-z_{-1})^2 - (z_{-2}-z_{-1})^2) / a 
\end{align}
with
\begin{equation}
a = (z_{-2}-z_{-1})(z_{-3}-z_{-1})(z_{-3}-z_{-2}).
\end{equation}

\subsubsection{Volume change}
Volume change is accounted for by adjusting in each time step the
distance between the points $dz$ (representing the volume) and
rescaling the concentration appropriately. As the concentration is
defined at the discretisation points, but the volume in between these
points, two differently located scaling factors will be used.

As the first step, the initial masses $m$ are computed both, at and
between the discretisation points. After solving the diffusion
equation, the mass fluxes of Li, $\dot m$, are calculated at
and between the points. This way, the exact masses are obtained in
each time step as 
\begin{equation}
m = m_{t-1} + \dot m\Delta t,
\end{equation}
with $t-1$ denoting the previous time step. The masses are used to
calculate the volume as 
\begin{equation}
V = m/\rho
\end{equation}
with the concentration-dependent density obtained from
equation (\ref{eqn:density:c}). The volumetric correction factor is then
simply defined as
\begin{equation}
f = V/V_{t-1}
\end{equation}
and is used to scale the distances between the discretisation points as
\begin{equation}
dz = f\cdot dz_{t-1}.
\end{equation}
Due to the fact that the amount of moles of Li stays constant during this
increase in volume, the molar concentration $c$ needs to be divided by
the same factor. As the concentration is defined at the discretisation
points, a similar factor is computed there. To stabilise the
solution, it is sometimes useful to perform the volume change operation
only every n$^\mathrm{th}$ iteration.

\subsection{Finite volume method}\label{a:diffusionNumericalFVM}
\subsubsection{Equation}
The diffusion equation
\begin{equation}
\frac{\partial c}{\partial t} = \nabla\cdot D\nabla c
\end{equation}
is discretised using the finite volume method as \cite{Ferziger2008,
  Versteeg2007,Schwarze2013,Moukalled2016}
\begin{equation}
\frac{\partial}{\partial t}\int_V cdV = \int_V\nabla\cdot\left(D\nabla c\right)dV.
\end{equation}
Using the implicit Euler method and the discretized form of the Gauss theorem, we obtain
\begin{equation}
\frac{(cV)^t - (cV)^{t-1}}{\Delta t} = \sum_f D_f \bi S_f\cdot(\nabla c)_f = \sum_f D_f |S_f|\frac{c_N - c_P}{d}
\end{equation}
with $V$ denoting the volume of the control volume, $D_f$ the
diffusivity on the faces, $\bi S_f$ the surface area vector, $d$ the
distance between two cell centres, $c_P$ the concentration in the
parent cell and $c_N$ the concentration in the neighbour cell. By
dividing the equation by the surface area and introducing the cell
height $dz$ we obtain
\begin{equation}
c^t = \sum_f D_f \frac{c_N - c_P}{d}\frac{\Delta t}{dz^t} + \frac{c^{t-1}dz^{t-1}}{dz^t}.
\end{equation}
Denoting the concentration in the lower cell by $c_{i-1}$ and the one
in the upper cell by $c_{i+1}$, $c_N$ is replaced. Further, we denote
the distance of the parent cell centre to the lower cell by $d_l$ and
the distance to the upper cell as $d_u$ and set $c^t = c_i$. We obtain
the discretised equation in the matrix-form
\begin{equation}
A\cdot c = b,
\end{equation}
as
\begin{equation}
c_{i-1}\left(D_f\frac{\Delta t}{d_ldz^t}\right) + c_i\left(-1
-\frac{D\Delta t}{d_l dz^t} - \frac{D_f\Delta t}{d_u dz^t}\right) +
c_{i+1}\left(\frac{D_f\Delta t}{d_u dz^t}\right) = -\frac{c^{t-1}dz^{t-1}}{dz^t}
\end{equation}
Assuming no volume change at small time steps, we can assume
$dz^t\approx dz^{t-1}$.

\subsubsection{Boundary conditions}
The boundary condition at the lower interface reads
\begin{equation}
\nabla c\cdot\bi n = 0 = -c_1 + c_0,
\end{equation}
which means basically that the flux over the lower interface needs to
be set to zero. This leads to a modification of the coefficients as
$A[0,0] = -1-\frac{D_f\Delta t}{d_u dz^t}$.
Similarly, the boundary condition on the top interface
\begin{equation}
\nabla c\cdot\bi n =\frac{j}{\nu_eFD}
\end{equation}
requires a constant (given) flux through the upper interface,
leading to a modified source term as
\begin{equation}
b[-1] = -\frac{c^{t-1}dz^{t-1}}{dz^t} - \frac{j\Delta t}{\nu_eFdz^t}
\end{equation}
and a modified diagonal coefficient as
\begin{equation}
A[-1,-1] = -1-\frac{D_f\Delta t}{d_l dz^t}.
\end{equation}

\subsubsection{Volume change}
Likewise to the finite difference method, the cell volume and
concentration are scaled in each time step to account for volume
change. However, here only one scaling factor is needed as
concentration and volume are both defined in the cell centre of the
control volumes.

\end{document}